\documentclass[preprint]{aastex}

\usepackage{graphicx}
\usepackage{float}
\usepackage{epstopdf}
\usepackage{multirow}
\usepackage{array}
\usepackage{booktabs}
\usepackage{natbib}
\begin{document}

\title{Optimization of NANOGrav's Time Allocation for Maximum Sensitivity to Single Sources}

\author{Brian Christy \altaffilmark{1}\altaffilmark{*}, Ryan Anella \altaffilmark{1}, Andrea Lommen \altaffilmark{1}, Lee Samuel Finn \altaffilmark{2}, Richard Camuccio \altaffilmark{1}, Emma Handzo \altaffilmark{1}}
\altaffiltext{1}{Franklin and Marshall College, Dept of Physics and Astronomy, Lancaster, PA, 17604}
\altaffiltext{2}{Dept of Physics, Dept of Astronomy and Astrophysics, The Pennsylvania State University, University Park, PA, 16802}
\altaffiltext{*}{email: brian.christy@fandm.edu}
\date{\centering \today \linebreak Submitted v2.0}

\begin{abstract}
Pulsar Timing Arrays (PTAs) are a collection of precisely timed millisecond pulsars (MSPs) that can search for gravitational waves (GWs) in the nanohertz frequency range by observing characteristic signatures in the timing residuals.  The sensitivity of a PTA depends on the direction of the propagating gravitational wave source, the timing accuracy of the pulsars, and the allocation of the available observing time. The goal of this paper is to determine the optimal time allocation strategy among the MSPs in the North American Nanohertz Observatory for Gravitational Waves (NANOGrav) for a single source of GW under a particular set of assumptions.  We consider both an isotropic distribution of sources across the sky and a specific source in the Virgo cluster.  This work improves on previous efforts by modeling the effect of intrinsic spin noise for each pulsar.  We find that, in general, the array is optimized by maximizing time spent on the best-timed pulsars, with sensitivity improvements typically ranging from a factor of 1.5 to 4.  
\end{abstract}

\section{Introduction}
The precision timing of pulses from millisecond pulsars (MSPs)
\citep{Backer82} is a valuable tool for studying astronomical objects.
These clocks in space can be used to make stringent tests on gravity
\citep{Kramer06}, study the interstellar medium between the pulsar and Earth \citep{You07}, and have even produced the indirect detection of gravitational waves (GWs) \citep{Hulse75,Taylor92}.  
Pulsar Timing Arrays (PTAs) aim to advance the GW effort with a direct detection.  By studying the small differences between the measured and expected arrival time of the pulses, PTAs can be used to search for the passage of a GW \citep{Foster90,Jenet05detect,Oslowski11}.  The frequency range, set by the cadence and time span of the observations, is in the nanohertz regime.  

The primary goal is to detect a correlated disturbance in the arrival times across several pulsars \citep{hellings:gw} that would result from the GW passing the Earth.  The correlation allows for the unambiguous distinction of the GW signal from possible sources of timing noise in the individual pulsars.  There are currently three collaborations that use PTAs as detectors for GWs: the North American Nanohertz Observatory for Gravitational Waves (NANOGrav) \citep{Jenet09}, the European Pulsar Timing Array (EPTA) \citep{Ferdman10}, and the Parkes Pulsar Timing Array (PPTA) \citep{Manchester13}. Together they make up the International Pulsar Timing Array (IPTA) \citep{Hobbs10}.

There are two types of GW signals that PTAs search for: single sources and the resulting confusion limit or stochastic background. The most plausible astrophysical source for these signals are supermassive black hole binaries (SMBHB) \citep{jb03,svc08,Sesana13}. A single source could be a bright SMBHB above the stochastic background or a burst source whose lifetime is shorter than the observation time of the array.  Burst sources might arise from the close encounter of a highly elliptical SMBHB or from cosmic strings \citep{Finn10,Siemens07,Ravi14}.  Further, a burst may not pass the Earth during observation but the long term effect of a permanent metric change could be detected in the PTA due to the long photon travel time, a so-called burst with memory\citep{Cordes12}.  The time it takes to detect any of these sources will depend, among other things, on the allocation of time amongst the pulsars in the array. 

Several questions must be answered before the optimization is done.  Is the goal to detect single sources or the stochastic background?  Is the goal to maximize the number of sources or the signal-to-noise ratio (SNR) of the detected sources?  What is the total amount of time available for allocation?  Are all pulsars visible for this total time? Is the goal simply to detect sources or be able to characterize their waveforms?  In this paper, we focus on searching for single sources.  Work on optimization for a stochastic background can be found in \cite{Lee12} and \cite{Siemens13}. We seek to optimize the total number of detectable sources rather than the SNR of a particular source.  We investigate the role of total observing time by presenting results for time allocations up to 10x the amount currently used in the most recent published NANOGrav dataset \citep{Demorest13}.  We assume that the amount of time pulsars are viewable ($\sim 2-12$ hours/day depending on the declination and telescope) is longer than the amount of time available to be allocated, hence no restrictions are made for how much time is given to each pulsar.  Finally, our goal is detection and not characterization.  Future work \citep{Koop} will investigate how these results change when the goal is to characterize the detected waveforms.  

We present a simple way to allocate the available observing time among
the pulsars in NANOGrav to maximize the volume sensitivity. This work
extends the results of \citet{Burt11}, who indicate that a PTA can be
optimized by allocating more time to the best pulsars. They
characterize the noise of an individual pulsar by the root mean square
(RMS) of the timing residuals.  Their model uses the relationship that
the RMS is inversely proportional to the square root of the integrated
observing time. This is based on the assumption that the pulsars in
the array are dominated by white noise.  Indeed, most NANOGrav
  pulsars exhibit this property \citep{Demorest13,Handzo,Perrodin}.
However, it is presumed that the residuals do have a red-noise
component \citep{Shannon10}, yielding a point in which longer
integration times do not improve the RMS.  In general, we
  introduce a noise floor, with red noise as the most plausible
  source.  In this way, it accounts for any excess low frequency power that may be present, such as effects intrinsic to the pulsar \citep{Hobbs06} or the interstellar medium (ISM) \citep{Hemberger08}.

As the current MSPs still exhibit decreasing RMS for longer
integration times, an accurate value for a noise floor is unavailable.  For this work, we investigate three noise floor models. The first model uses a constant 10 ns noise floor for each pulsar in the PTA. The second model uses $80\%$ of the current RMS of the timing residuals for each pulsar as the noise floor. For the final model, we run 1000 simulations of the detector with a noise floor of each pulsar drawn from a uniform distribution between these two extremes.  We then examine the average time allocations and distributions for each pulsar from these simulated arrays.

A common assumption in many optimization studies is that the sources are distributed isotropically.  In this work, we investigate whether the optimization would change if a particular direction is favored.  Recent work by \citet{Simon14} looks at the distribution of galaxies in the local universe to suggest that nearby galaxy clusters are the most promising locations for a bright single source.  Therefore, we also apply our optimization scheme to a single GW source emanating from Virgo as a particular example.

The outline of this paper is as follows: \S 2 identifies the figures of merit to be optimized, drawing heavily from \citet{Burt11}, and expands on these results by including the noise floor. \S 3 introduces the three different noise floor models and presents the timing allocation results when these models are optimized. \S 4 presents the increase in sensitivity and discusses the results. \S 5 concludes by placing these results in context of the wider studies and describes potential future investigations.

\section{Sensitivity in a PTA}
Our goal is to maximize the total number of single sources a PTA can detect.  In the first optimization, we assume an isotropic and homogenous distribution of sources and maximize the volume of space in which a source of a particular strength would be detectable.  This follows the same assumption of \citet{Burt11}.  We start by considering a fiducial source propagating in the $\hat{k}$ direction.  The time-averaged power signal to noise ratio $\overline{\rho^2(\hat{k})}$ is (see Equation 16 of \cite{Burt11}):
\begin{equation}
\label{eqn:rho_squared}
\overline{\rho^2(\hat{k})}=\frac{A^2}{4}\sum_{j=1}^{n_p}(\frac{1-\hat{k}\cdot\hat{n}_j}{\sigma_j})^2
\end{equation}
Here, $A$ is the amplitude of the propagating GW, $\hat{k}$ is the
direction it is propagating, $\hat{n}$ is the unit vector pointing to
pulsar $j$, and $\sigma_j$ is the RMS of the timing residuals for the
$j$th pulsar.  In deriving this result, the response to gravitational
waves is split into an Earth term  and pulsar term (see, e.g.,
\cite{Jenet04}), where the pulsar term is delayed by the difference
between when the GW passes Earth and the information that the GW
passes the pulsar reaches Earth.  For nominal cases, this is on the
order of the light travel time from the pulsar ($\sim 1,000$ years)
but ranges from 0 to twice the light travel time.  

In Equation \ref{eqn:rho_squared}, only the Earth term 
  contributes to the SNR.  For burst sources that are shorter than the length of the
  data, the pulsar term delay means it will typically not appear in the
  data.  For continuous sources that last many centuries, there will
  be a pulsar term in the data that affects the SNR.  However, as
argued in \citet{Burt11}, including this pulsar term requires
knowledge of the distance to the pulsar.  It enters into the residual 
response in the form of $\cos{\omega(t-(1-\hat{k}\cdot\hat{n})L)}$,
where $\omega$ is the frequency of the source and L is the pulsar distance.
In the absence of an exact value for L, one must 
average over the range of possible values.  Typical uncertainties in
pulsar distances are $\sim$ 10\% \citep{Cordes02}, resulting in typical
ranges for $\omega(t-(1-\hat{k}\cdot\hat{n})L)$ that are many times
2$\pi$ for the frequencies of interest.  Therefore, when averaging
over the uncertainty in the pulsar distance, the result to the SNR in
Equation \ref{eqn:rho_squared} will be 0.  In
other words, the lack of knowledge for the pulsar distance means that
one cannot calculate its contribution to this optimization.

An important caveat for both burst and continuous sources is
when $(1-\hat{k}\cdot\hat{n})$ approaches 0.  For bursts, this means 
the delay in the pulsar term approaches 0, while for continous
sources, it means the uncertainty from L becomes inconsequential.  
The extreme is when
the source is in line with the pulsar, and the pulsar term effectively
cancels the Earth term and the signal is 0.  For the remainder of this
work, we will assume the source is far enough separated in the sky
from the pulsar such that for burst sources the pulsar term is not in
the data and for continous sources the effect of averaging over the
pulsar distance uncertainty leads to 0 for its contribution to the
sensitivity.  As an example, for a burst source with duration $\Delta
T$ and pulsar distance L, the pulsar term will not affect the Earth
term as long as the sky separation is larger than \citep{Finn10}:

\begin{equation}
\label{eqn:burstsep}
\theta (\rm{degrees}) \approx 2.5\times(\frac{\Delta
  T}{\rm{year}}\frac{\rm{kpc}}{L})^{1/2}
\end{equation}

\citet{Burt11} demonstrate that for a given source strength, the volume which it will be detectable scales as $\overline{\rho^2(\hat{k})}^{3/2}$.  This follows directly from the fact that the amplitude $A$ of the GW source is inversely proportional to the source distance.  For example, take a fiducial source that has an SNR ($\sqrt{\overline{\rho^2(\hat{k})}}$) just above the threshold of detection.  If one cuts the noise in half, the detectable amplitude is cut in half as well.  This doubles the distance out to which the fiducial source can be located to still be detected, which in turn increases the detectable volume by a factor of 8.  

We call this the volume sensitivity and, for a given direction, it is defined as:
\begin{equation}
\label{eqn:nu_k}
\nu(\hat{k}) = \frac{\overline{\rho^2(\hat{k})}^{3/2}}{\overline{\rho_{\rm{ref}}^2(\hat{k})}^{3/2}}
\end{equation}
Here we compare the volume sensitivity to a reference PTA in which the timing is distributed equally among all pulsars.  Equally timing all pulsars is the typical goal of current timing campaigns, and this comparison will give the expected improvement in sensitivity while making a specific amplitude unnecessary.

While Equation \ref{eqn:nu_k} gives a sensitivity for a particular source direction, we can characterize the overall sensitivity of the PTA by integrating over the entire sky.  In practice, this is done by summing up over discrete pixels in the sky:
\begin{equation}
\label{eqn:nu_overall}
\nu_{\mathrm{overall}}=\frac{\displaystyle\sum_{i=1}^n\overline{\rho^2(\hat{k}_i)}^{3/2}}{\displaystyle\sum_{i=1}^n\overline{\rho_{\mathrm{ref}}^2(\hat{k}_i)}^{3/2}}
\end{equation}
Here, $n$ is the the number of pixels in the integration.  We use \texttt{HEALPIX} \footnote{http://healpix.jpl.nasa.gov} to divide the sky into 3072 equal-area pixels.  

We note that without the summation over the sky, the maximum of $\nu(\hat{k})$ is the same as the maxima of $\overline{\rho^2(\hat{k})}$ and $\sqrt{\overline{\rho^2(\hat{k})}}$, so the goal of maximizing the volume of detectable sources and the SNR of a source is equivalent.  Looking over the sky, however, these goals may diverge due to the summation.  Thus, while our goal is to maximize the number of detectable sources, when we consider a single direction, this is equivalent to maximizing the SNR.

A primary task of this work is to characterize $\sigma_j$ as a function of integration time.  \cite{Burt11} assumed white noise which leads to the following relationship:
\begin{equation}
\label{eqn:whitenoise}
\sigma_j =\sigma_{j0}\sqrt{\frac{T_{j0}}{T_{j,\rm{obs}}}} = \sigma_{j0}\sqrt{\frac{T_{j0}}{t_{j,\rm{frac}}T_{\rm{Tot}}}}
\end{equation}
Here, $T_{j,\rm{obs}}$ is the amount of observation time for each pulsar that is integrated to produce a timing residual and $\sigma_{j0},\,T_{j0}$ represent the input values from \citet{Demorest13}, as described in \S \ref{sec:res}.  In general, we will use the subscript 0 to represent values of the \citet{Demorest13} dataset.  As more time in an observation is dedicated to a pulsar, $\sigma$ will decrease following a $T^{-1/2}$ relationship.  The second half of Equation \ref{eqn:whitenoise} is re-written in terms of the fractional amount of time spent on the $j$th pulsar ($t_{j,\rm{frac}}$) versus the total amount of time available ($T_{\rm{Tot}}$).  

We introduce a time independent noise floor $\sigma_r$ to account for any noise process that does not decrease with integration time, such as red noise:
\begin{equation}
\label{eqn:rednoise1}
\sigma_j^2(t_{j,\rm{frac}}) = \sigma_r^2+\frac{\sigma_{j0w}^2}{t_{j,\rm{frac}}T_{\rm{Tot}}/T_{j0}}
\end{equation}
Here, $\sigma_{j0w}$ represents the portion of the input RMS that is attributable to white noise.  To ensure that $\sigma_j(T_{j0}/T_{\rm{Tot}})=\sigma_{j0}$, we set $\sigma_{j0w}^2=\sigma_{j0}^2-\sigma_r^2$. 

We make the simplifying assumption that, for the input data values, the observation time is roughly equal across each pulsar.  Therefore, $N_PT_{j0} = T_{\rm{Tot}0}$ where $N_P$ is the number of pulsars in the data and $T_{\rm{Tot0}}$ is the total amount of time available for the \citet{Demorest13} dataset.  Then, Equation \ref{eqn:rednoise1} becomes:
\begin{equation}
\label{eqn:rednoise2}
\sigma_j^2(t_{j,\rm{frac}}) = \sigma_r^2+\frac{\sigma_{j0w}^2}{t_{j,\rm{frac}}N_P(T_{\rm{Tot}}/T_{\rm{Tot}0})}
\end{equation}

To further simplify, we assume each pulsar is observed by a
  single telescope, allowing for a 
single noise model per pulsar.  In the PPTA and
NANOGrav, pulsars are generally observed with a single telescope, with
J1713+0747 in NANOGrav being the lone exception.  However, in the EPTA
pulsars are regularly observed by many telescopes so a given pulsar
will have different white noise levels depending on the telescope.
This will also be true for any IPTA dataset that combines all of
these.  A future goal is to extend this work in order to optimize the
time on each pulsar and on each telescope, taking into account
telescope differences.

For the reference PTA values, we assume the allocation is equal across all pulsars, hence $t_{j,\rm{frac}}=1/N_P$.  Therefore, the equivalent form of $\sigma_j$ is:
\begin{equation}
\label{eqn:rednoiseref}
\sigma_{j,\rm{ref}}^2 = \sigma_r^2+\frac{\sigma_{j0w}^2}{(T_{\rm{Tot}}/T_{\rm{Tot}0})}
\end{equation}

Using this functional form of $\sigma$, we seek to find the values of the array $t_{j,\rm{frac}}$ that maximize either $\nu(\hat{k})$ or $\nu_{\rm{overall}}$.  We constrain these results so that $\sum_{j=1}^{N_p}(t_{j,\rm{frac}})=1$ and each $t_{j,\rm{frac}}$ is non-negative.  We also investigate the effect of an increase in the total observing time available, with values of $T_{\rm{Tot}}/T_{\rm{Tot}0} = 1,5,10$.  This becomes necessary only with the inclusion of the constant noise floor, since the fraction $T_{\rm{Tot}}/T_{\rm{Tot}0}$ would cancel out of Equation \ref{eqn:nu_k} and \ref{eqn:nu_overall} in the case where $\sigma_r = 0$ (when substituting in Equation \ref{eqn:whitenoise}), producing equivalent results for any value.  

When considering an increased time allocation, one must be careful to avoid equating this to simply increasing the time span of the dataset.  An important assumption in this work is a constant red noise floor.  However, the red noise floor is a function of time span.  As the dataset covers more years and decreases the low frequency cutoff ($\sim 1/T_{\rm{span}}$), the red noise floor will increase \citep{Cordes13}.  In this paper, we instead consider increasing the allocation time per observation, or integration time.  We do note, however, that at the typical frequencies for continuous single sources or bursts, the noise is still dominated by white processes \citep{Cutler13}.  Therefore, one could still benefit from a longer time span by fitting out the low frequency power.  Averaging down the resulting data could be another way of producing a larger factor of available integration time.  The main drawback of this method is that it eliminates the sensitivity to lower frequencies where one expects a stronger signal in both the stochastic background and the memory effect of bursts.

Note that when we calculate $\nu(\hat{k})$ and $\nu_{\rm{overall}}$ for larger values of $T_{\rm{Tot}}/T_{\rm{Tot}0}$ (e.g. 5 or 10), we use the same ratio for both the optimized and reference array.

\section{Results}\label{sec:res}
\begin{table}
\caption{Information on individual pulsars used in the optimization}
\begin{center}
    \begin{tabular}{|c|c|c|}
    \toprule
    & \citet{Demorest13} &\\
    Pulsar &$\sigma_{j0}$ ($\mu$s)& $(1-\hat{k}_{\rm{Virgo}} \cdot \hat{n})^2$\\
    \hline
    \hline
J0030+0451 & 0.15 & 0.00 \\ 
\hline
J0613-0200 & 0.18 & 0.87 \\ 
\hline
J1012+5307 & 0.28 & 2.77 \\ 
\hline
J1455-3330 & 0.79 & 2.34 \\ 
\hline
J1600-3053 & 0.16 & 1.93 \\ 
\hline
J1640+2224 & 0.41 & 2.22 \\ 
\hline
J1643-1224 & 1.47 & 1.88 \\ 
\hline
J1713+0747 & 0.03 & 1.79 \\ 
\hline
J1744-1134 & 0.20 & 1.29 \\ 
\hline
J1853+1308 & 0.26 & 0.89 \\ 
\hline
B1855+09 & 0.11 & 0.84 \\ 
\hline
J1909-3744 & 0.04 & 0.53 \\ 
\hline
J1910+1256 & 0.71 & 0.76 \\ 
\hline
J1918-0642 & 0.20 & 0.58 \\ 
\hline
B1953+29 & 1.44 & 0.63 \\ 
\hline
J2145-0750 & 0.20 & 0.06 \\ 
\hline
J2317+1439 & 0.25 & 0.02 \\ 
 \bottomrule
\end{tabular}
\end{center}
\label{tab:PulsarInfo}
\end{table}
 For this work, we focus on the results of \citet{Demorest13}, which describes a timing campaign for 17 MSPs over $\sim\,$5 years.  Each pulsar received approximately equal amount of observing time at typical cadences of 4 to 6 weeks.  They were observed either at the 100-meter Green Bank Telescope or the 305-meter Arecibo Telescope.  A typical observation session included observing at two widely spaced frequencies to allow for correction of the dispersion measure (DM).  At Arecibo, these observations were typically spaced about an hour apart while at the GBT, the separation was up to a week.  Each pulsar was timed at each frequency for $\sim$15-45 minutes, with shorter integration times at Arecibo.  We take the input $\sigma_{j0}$ to be the reported (epoch-averaged) RMS values in \citet{Demorest13}.  These values range from $\sim$0.03$\mu$s to $\sim$1.5$\mu$s and are listed in Table \ref{tab:PulsarInfo}.  In all cases, we take the noise floor to be less than the reference $\sigma_{j0}$.  

We consider two optimization schemes for each of three noise floor models.  The first maximizes $\nu_{\rm{overall}}$ (Equation \ref{eqn:nu_overall}) to determine the time allocation $t_{j,\rm{frac}}$ for each pulsar that produces the largest volume of space for detectable sources across the whole sky.  The second maximizes $\nu(\hat{k})$ (Equation \ref{eqn:nu_k}) in the direction of the Virgo cluster.  The following plots demonstrate the resulting vectors $t_{j,\rm{frac}}$ for each noise model and the three different values of $T_{\rm{Tot}}/T_{\rm{Tot0}}$.  

The primary difference between between maximizing $\nu_{\rm{overall}}$ and $\nu(\hat{k})$ is where the pulsar is located relative to the Virgo cluster.  To help see the impact, the final column of Table \ref{tab:PulsarInfo} displays the numerical value of the directional term that enters into $\nu(\hat{k})$ (Equation \ref{eqn:nu_k}).  The larger this value, the more sensitive the pulsar is to a GW source from the Virgo cluster.

The three noise floor models are described below.

{\it Constant Noise Floor}: In this model we set the noise floor value to be a constant 10ns for each pulsar.  This is simplistic and likely optimistic, yet it provides a chance to gain intuition for the optimization.  The results are presented in Table \ref{fig:const_nf}.  The columns show the distribution of time that should be applied to the array in the cases of optimizing for the full sky and Virgo.  The rows represent different values of total time.

{\it 80\% Noise Floor Value for Each Pulsar}: In this model we set the noise floor to be 80\% of the measured RMS for each pulsar.  The input RMS values ($\sigma_{j0}$) are given in the second column of Table \ref{tab:PulsarInfo}.  This represents a pessimistic scenario and acts as a compliment to the constant 10ns noise floor.  The results are presented in Table \ref{fig:perc_nf}

{\it Noise Floor Drawn from Random}: The goal of this model is to present a more realistic situation where each noise floor is different while acknowledging our current ignorance of the value.  We perform 1000 iterations of the optimization where, for each case, the noise floor of each pulsar is drawn from a uniform distribution between 10ns and 80\% of the current RMS.  The time allocations for each iteration are then averaged down to a single value for each pulsar.  The results are shown in Table \ref{fig:rand_nf}.

\begin{table}
  \begin{center}
   \caption{Time allocation $t_{j,\rm{frac}}$ for a constant 10 ns
     noise floor.  Color version available online.} \begin{tabular}{|>{\centering\arraybackslash}m{0.16\textwidth}|>{\centering\arraybackslash}m{0.37\textwidth}|>{\centering\arraybackslash}m{0.37\textwidth}|}
        \toprule
        Time Allocation & Isotropic & GW from Virgo\\
        ($T_{\rm{Tot}}/T_{\rm{Tot0}}$) & & \\
        \midrule
        1x &
        \includegraphics[width=0.35\textwidth]{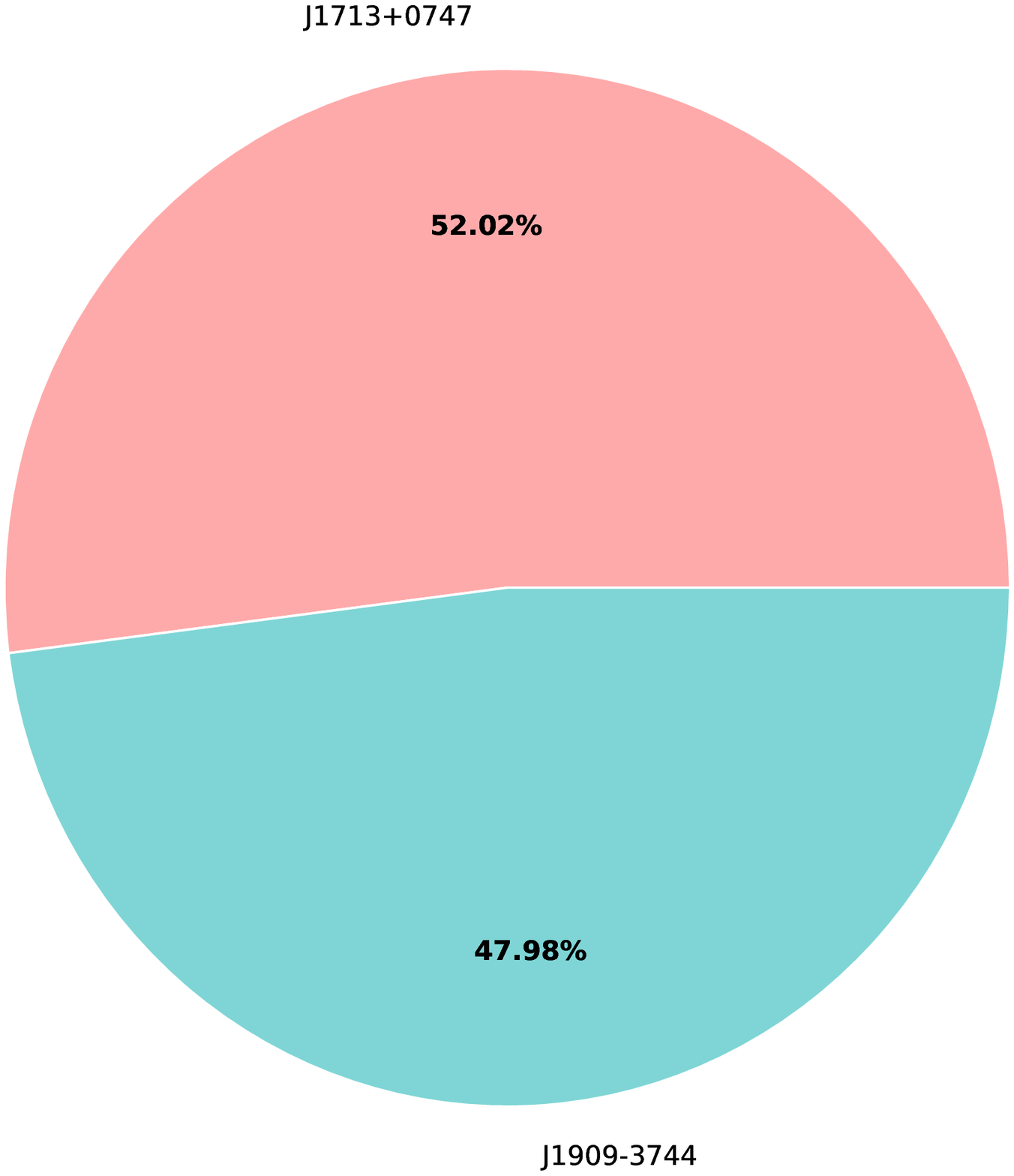} &
        \includegraphics[width=0.35\textwidth]{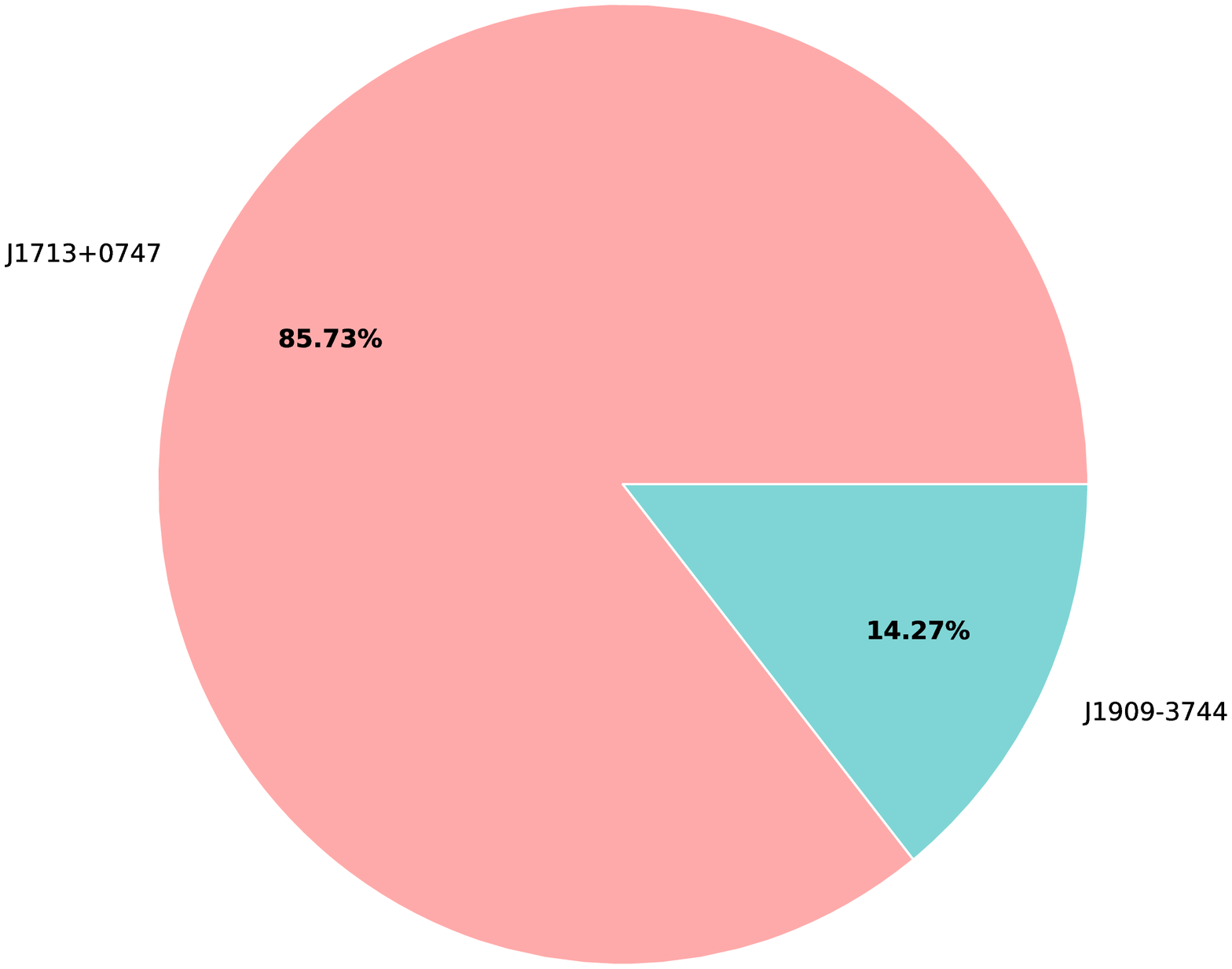} \\
        \midrule
        5x &
        \includegraphics[width=0.35\textwidth]{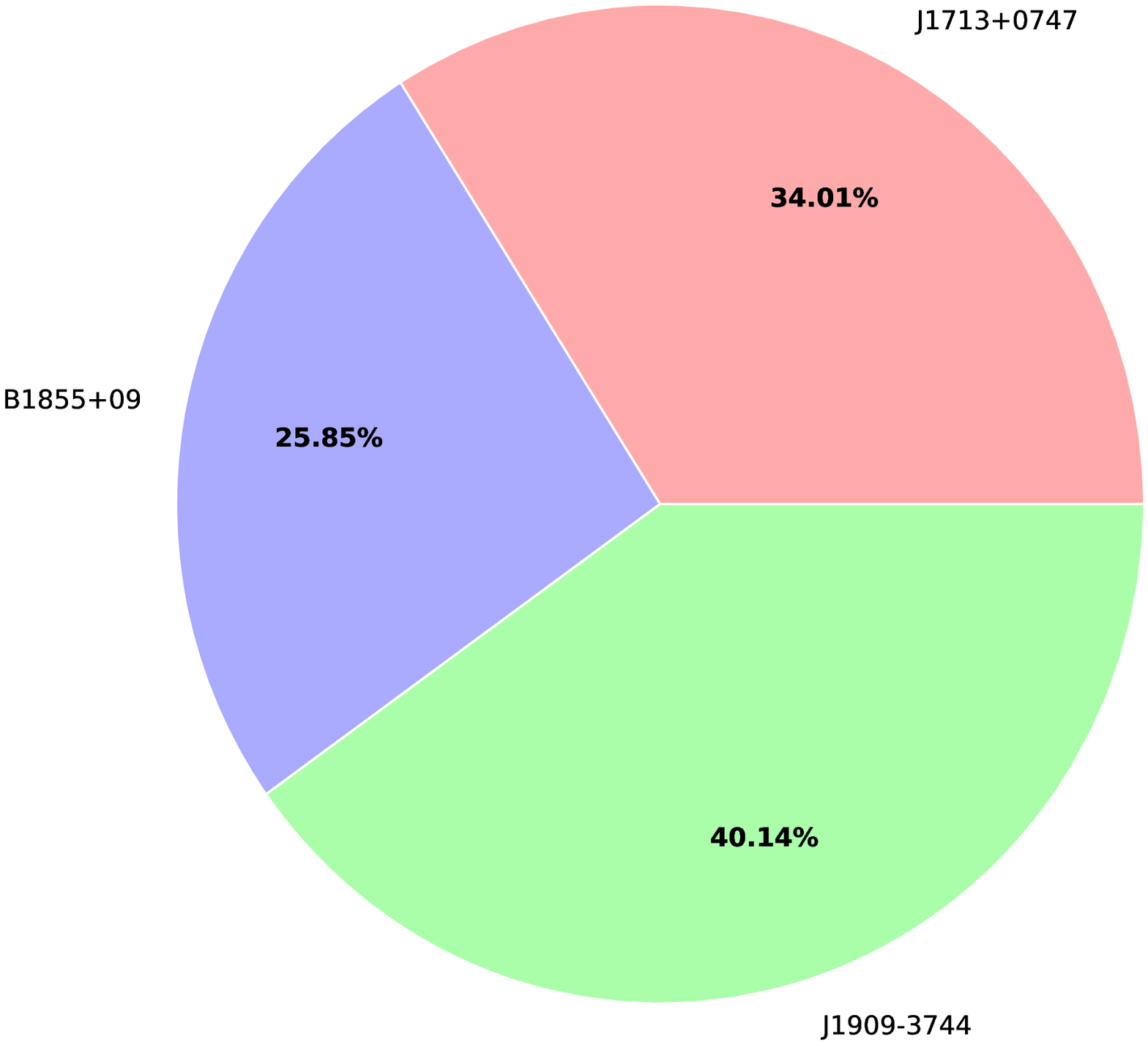} &
        \includegraphics[width=0.35\textwidth]{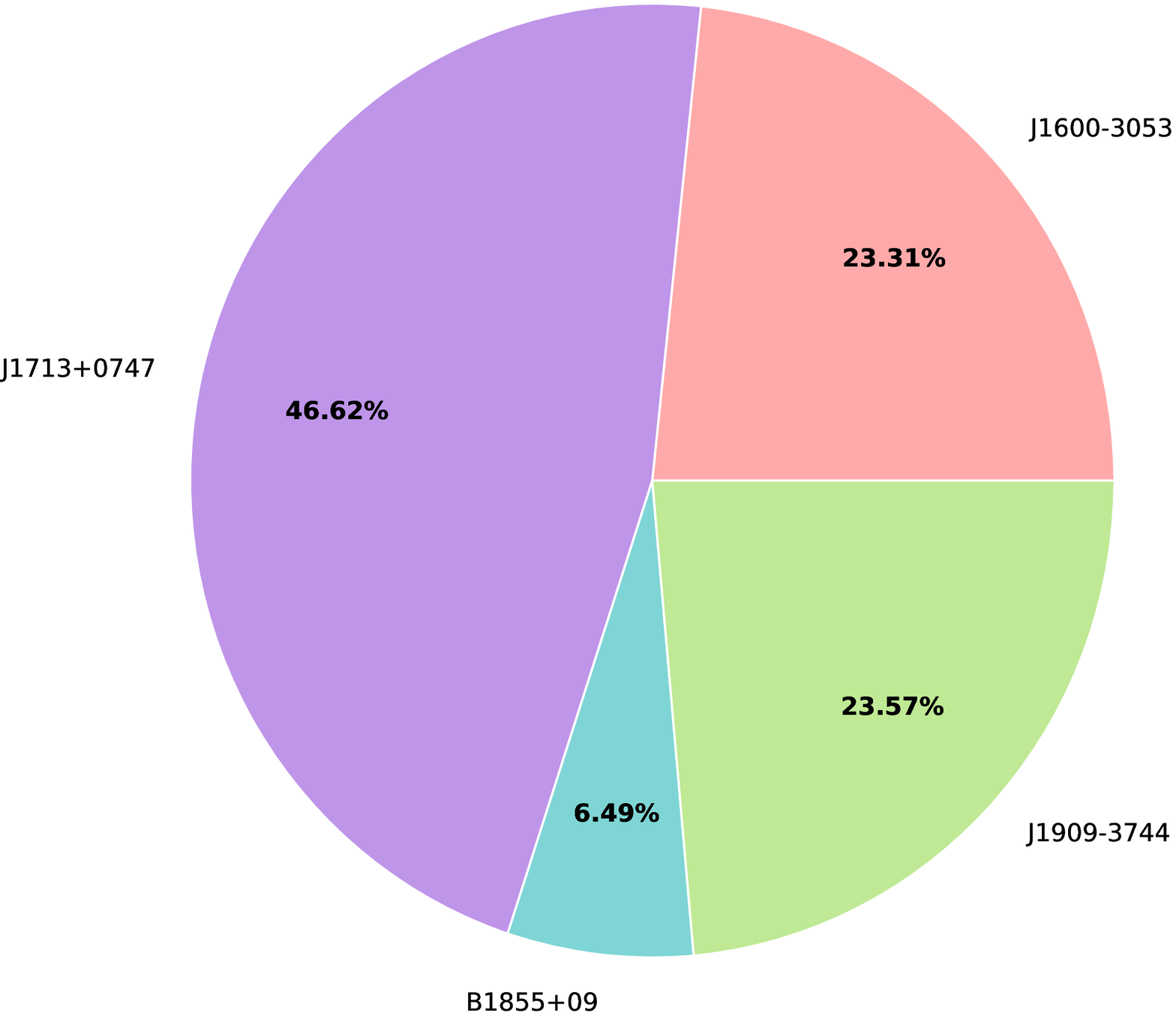} \\
        \midrule
        10x &
        \includegraphics[width=0.35\textwidth]{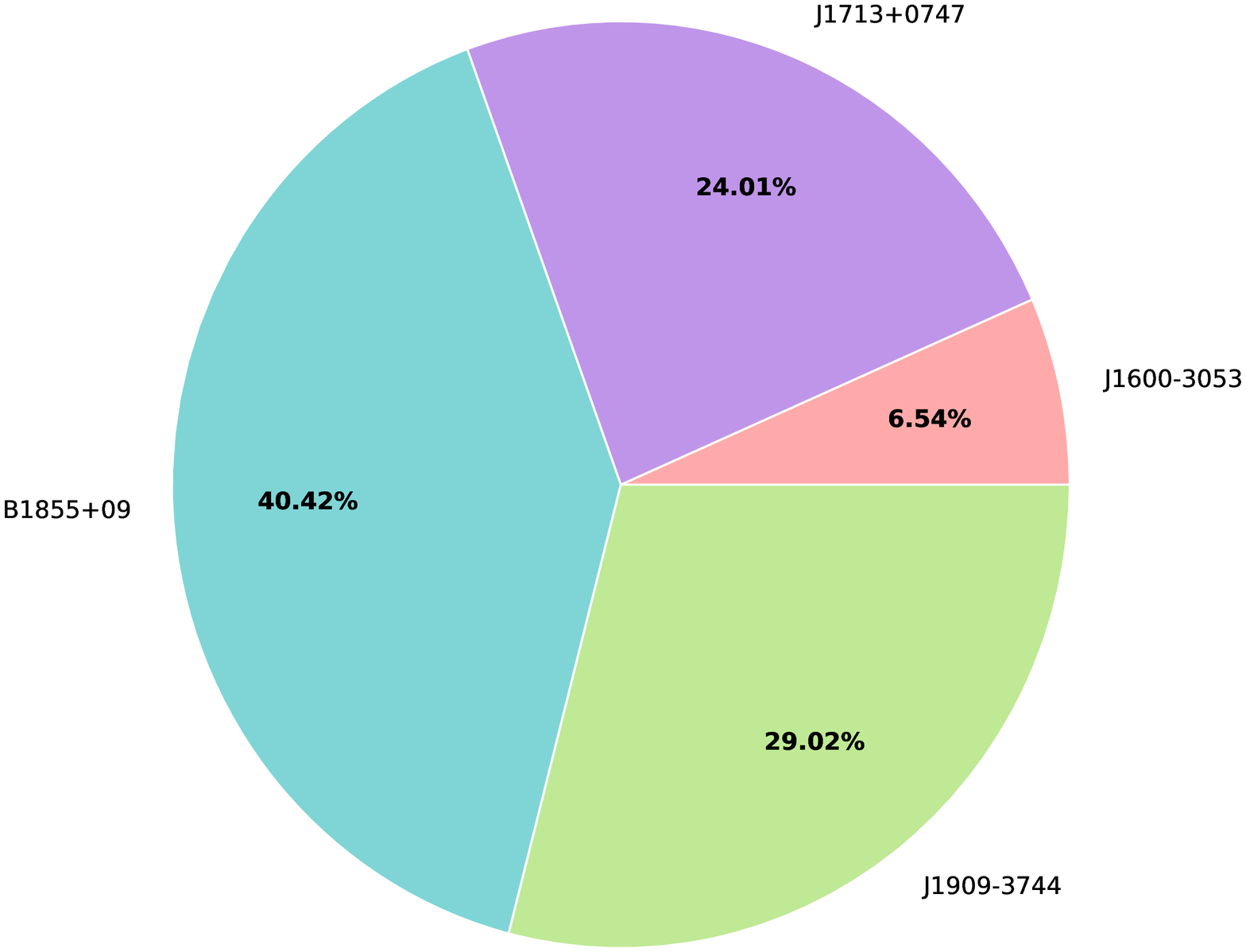}&
        \includegraphics[width=0.35\textwidth]{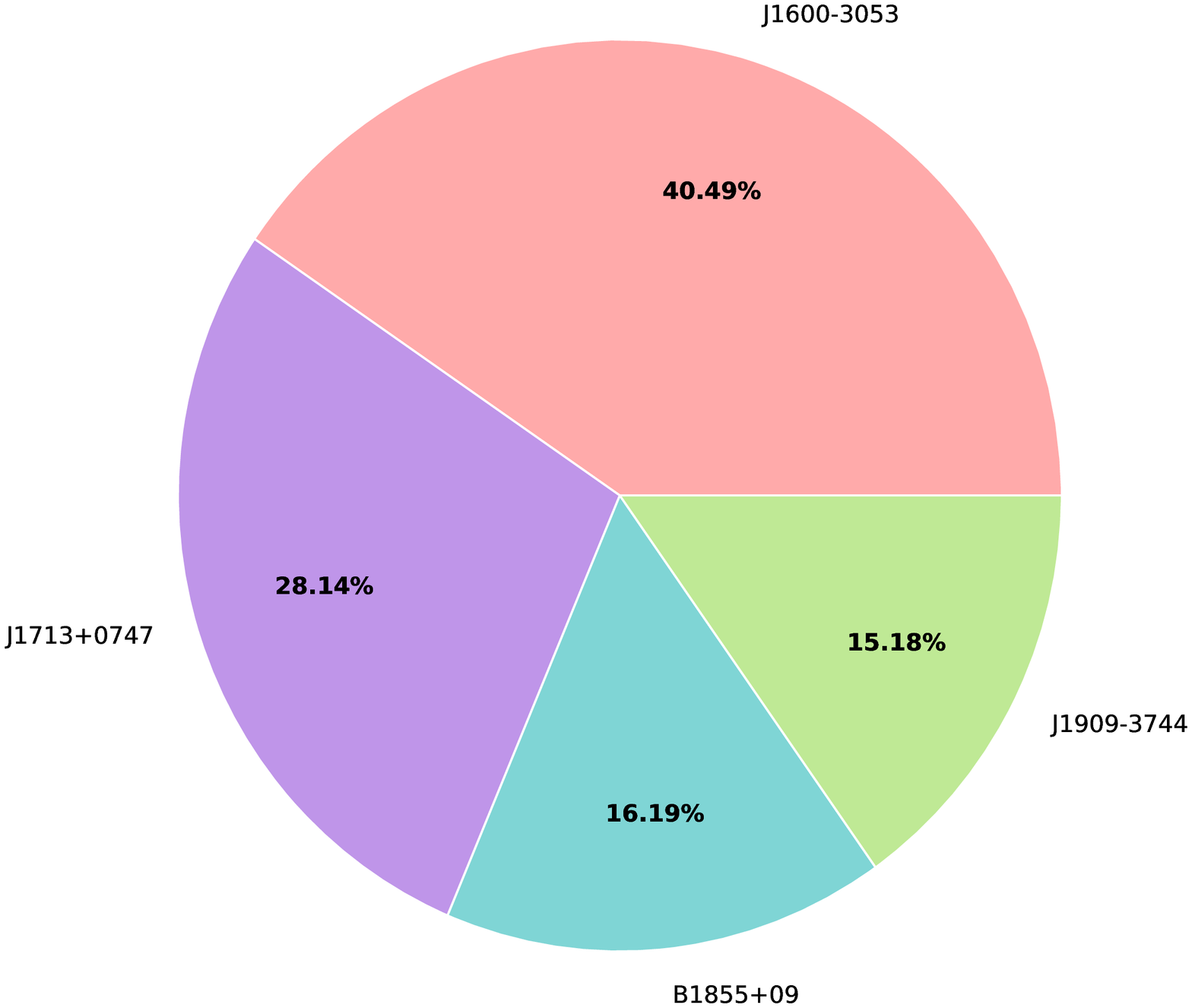}\\
        \bottomrule
        \end{tabular}
          \label{fig:const_nf}
    \end{center}
\end{table}

\begin{table}
  \begin{center}
   \caption{Time allocation $t_{j,\rm{frac}}$ for a noise floor 80\% the current level.  Color version available online.} \begin{tabular}{|>{\centering\arraybackslash}m{0.16\textwidth}|>{\centering\arraybackslash}m{0.37\textwidth}|>{\centering\arraybackslash}m{0.37\textwidth}|}
        \toprule
        Time Allocation & Isotropic & GW from Virgo\\
        ($T_{\rm{Tot}}/T_{\rm{Tot0}}$) & & \\
        \midrule
        1x &
        \includegraphics[width=0.35\textwidth]{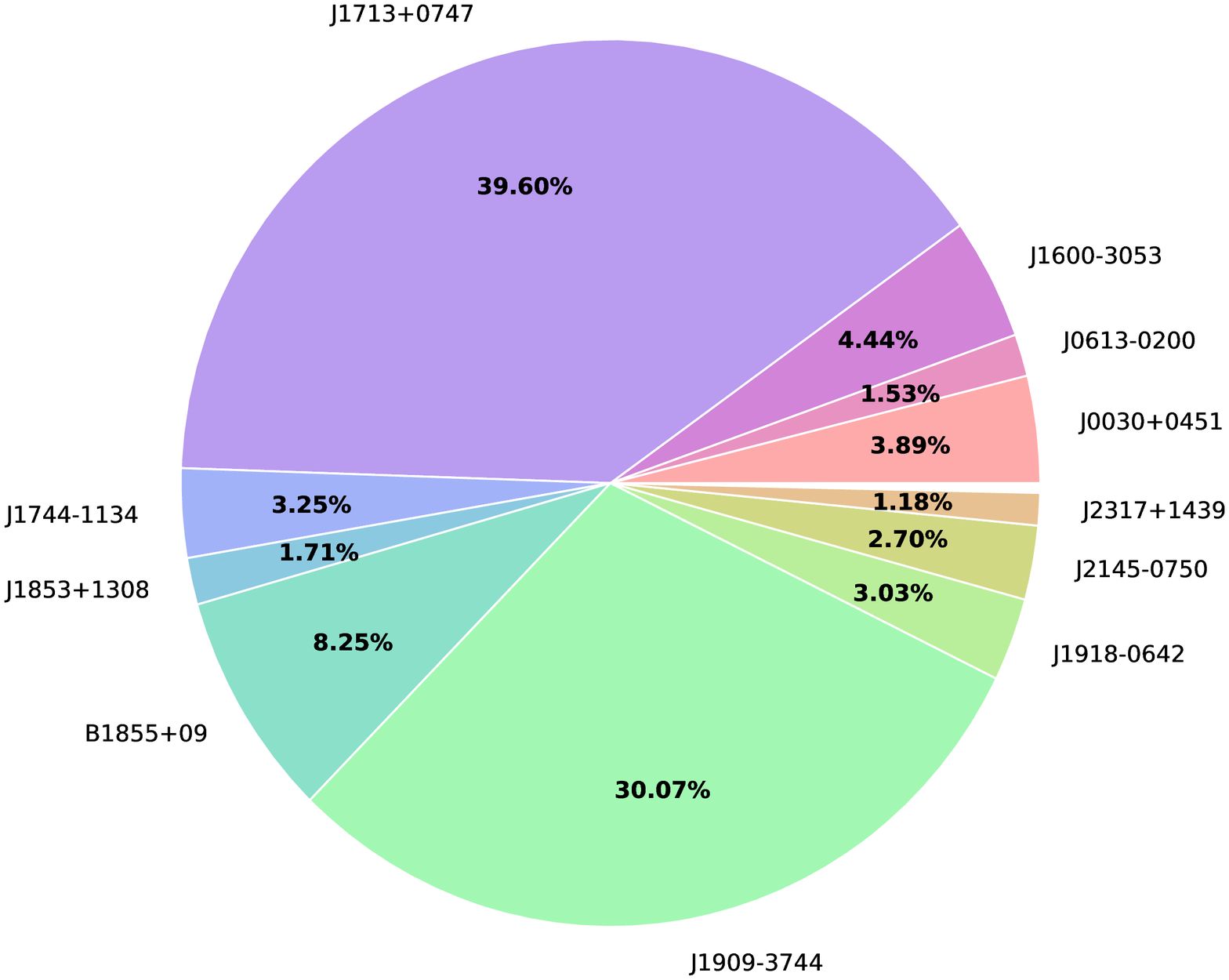} &
        \includegraphics[width=0.35\textwidth]{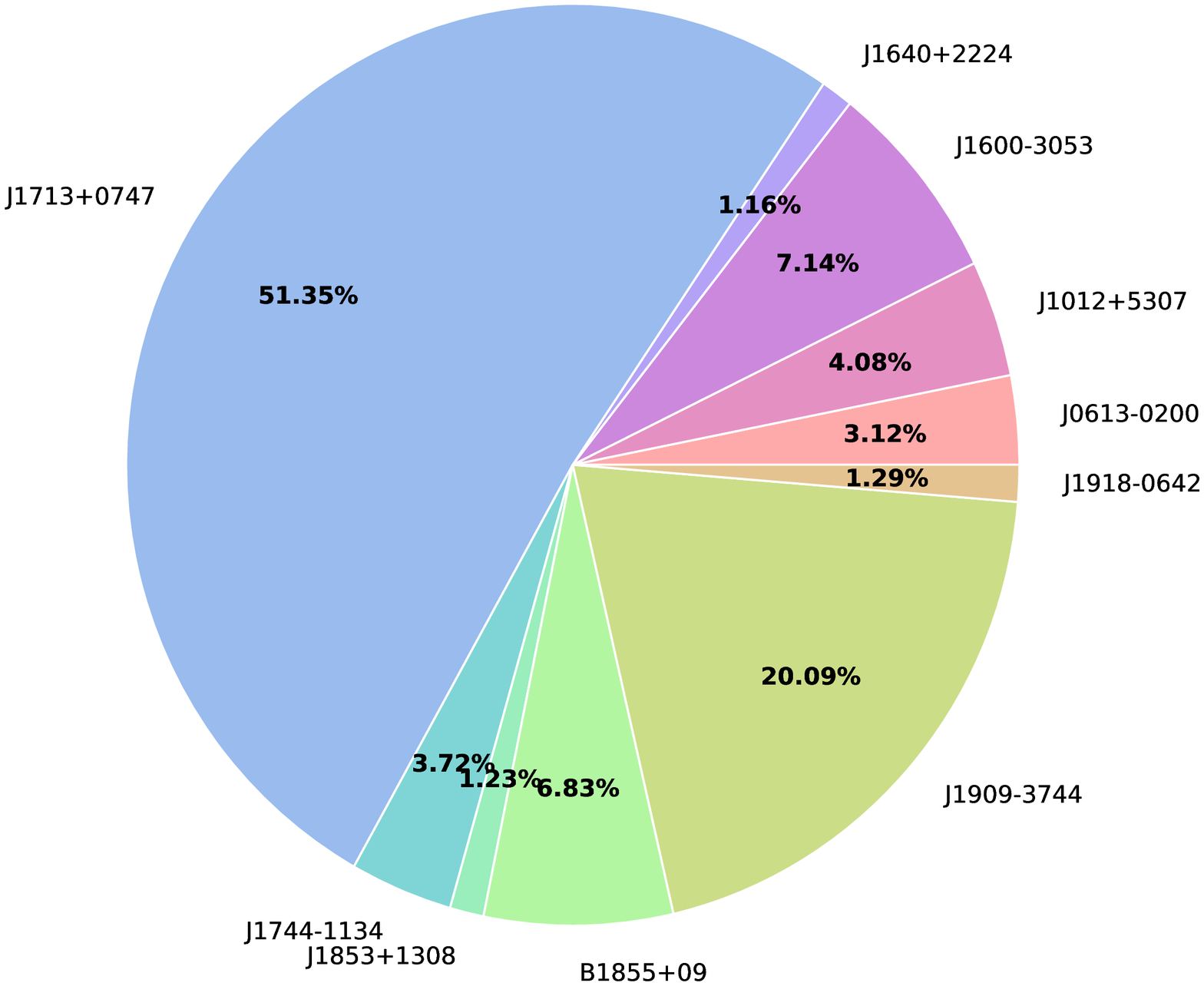} \\
        \midrule
        5x &
        \includegraphics[width=0.35\textwidth]{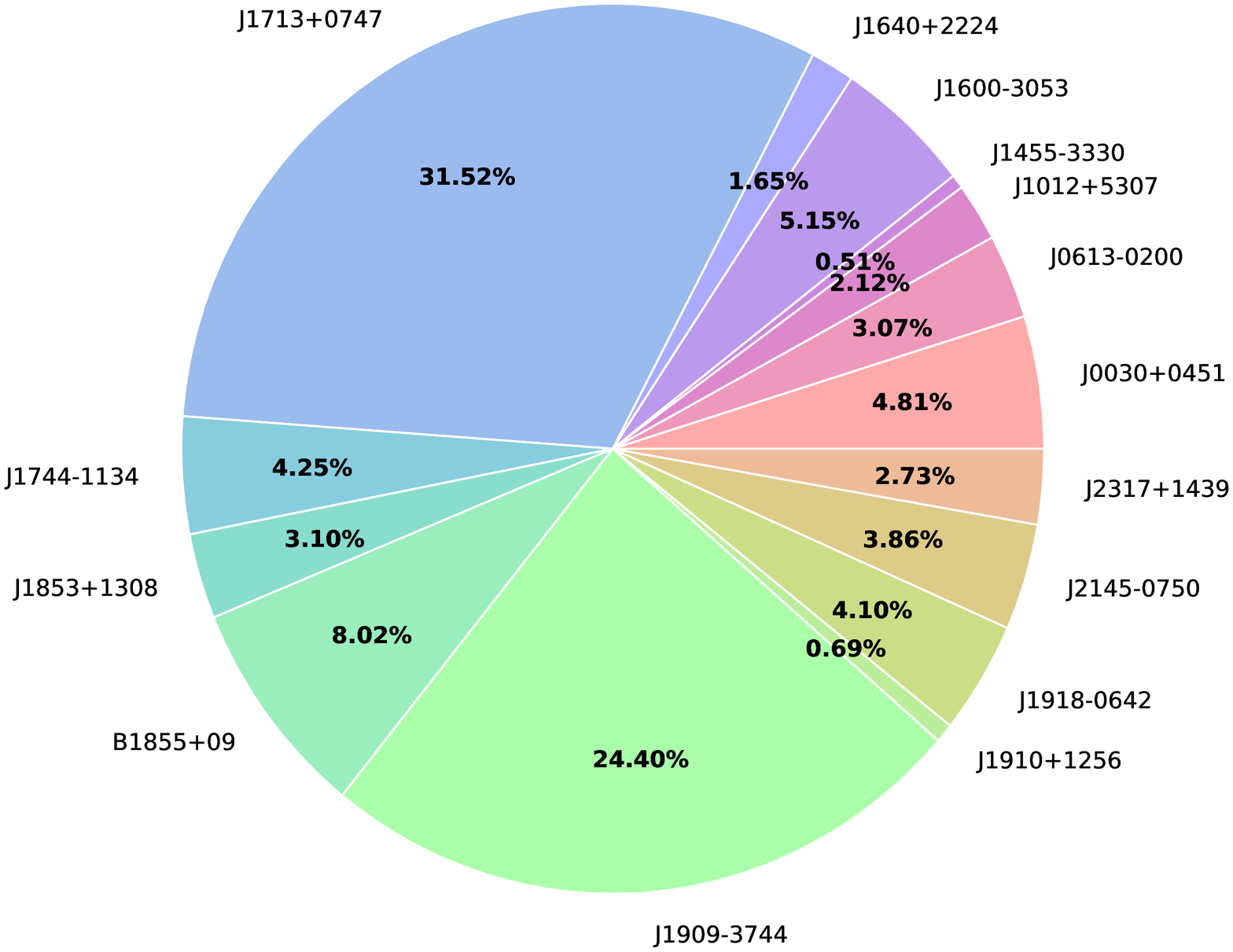} &
        \includegraphics[width=0.35\textwidth]{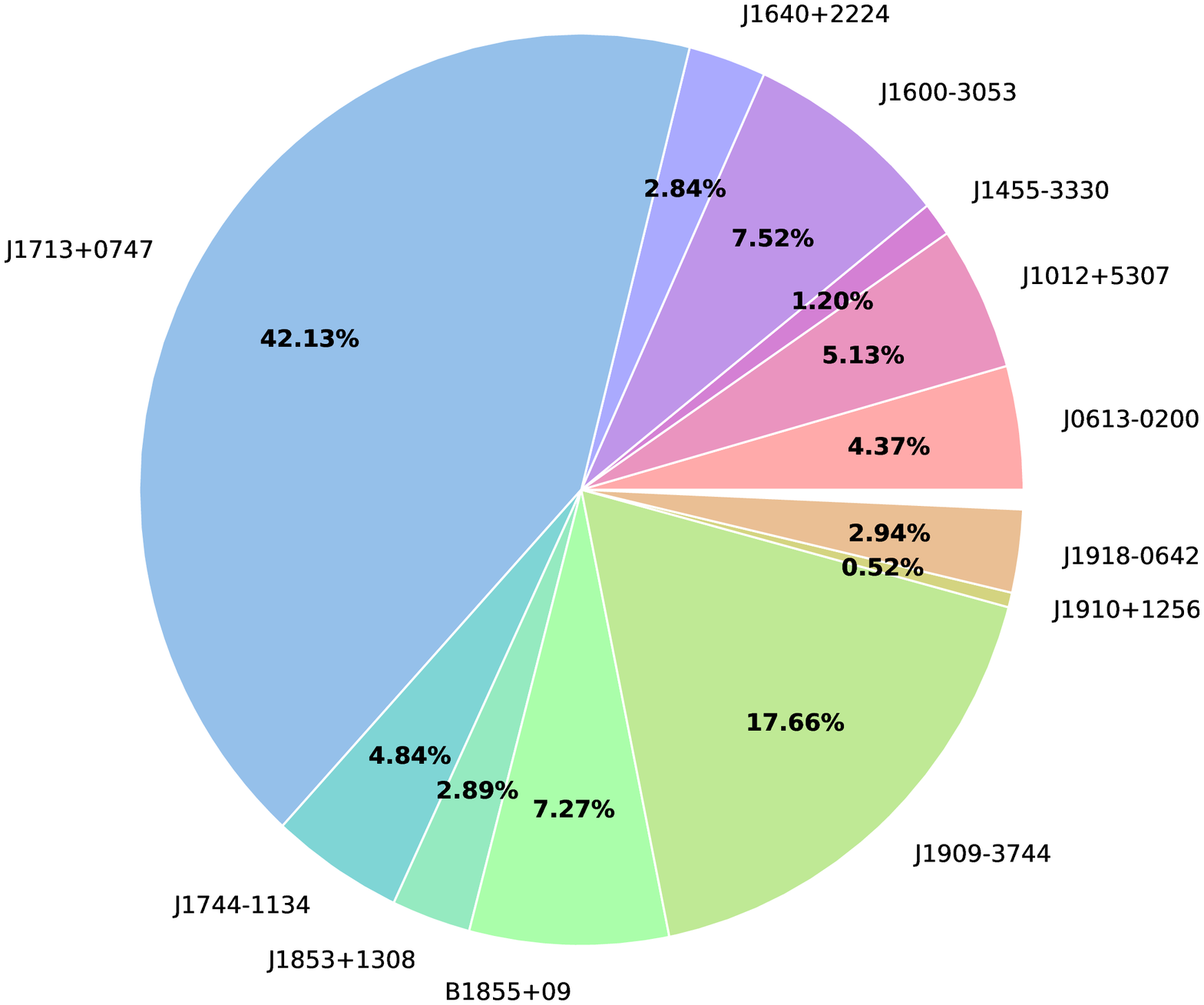} \\
        \midrule
        10x &
        \includegraphics[width=0.35\textwidth]{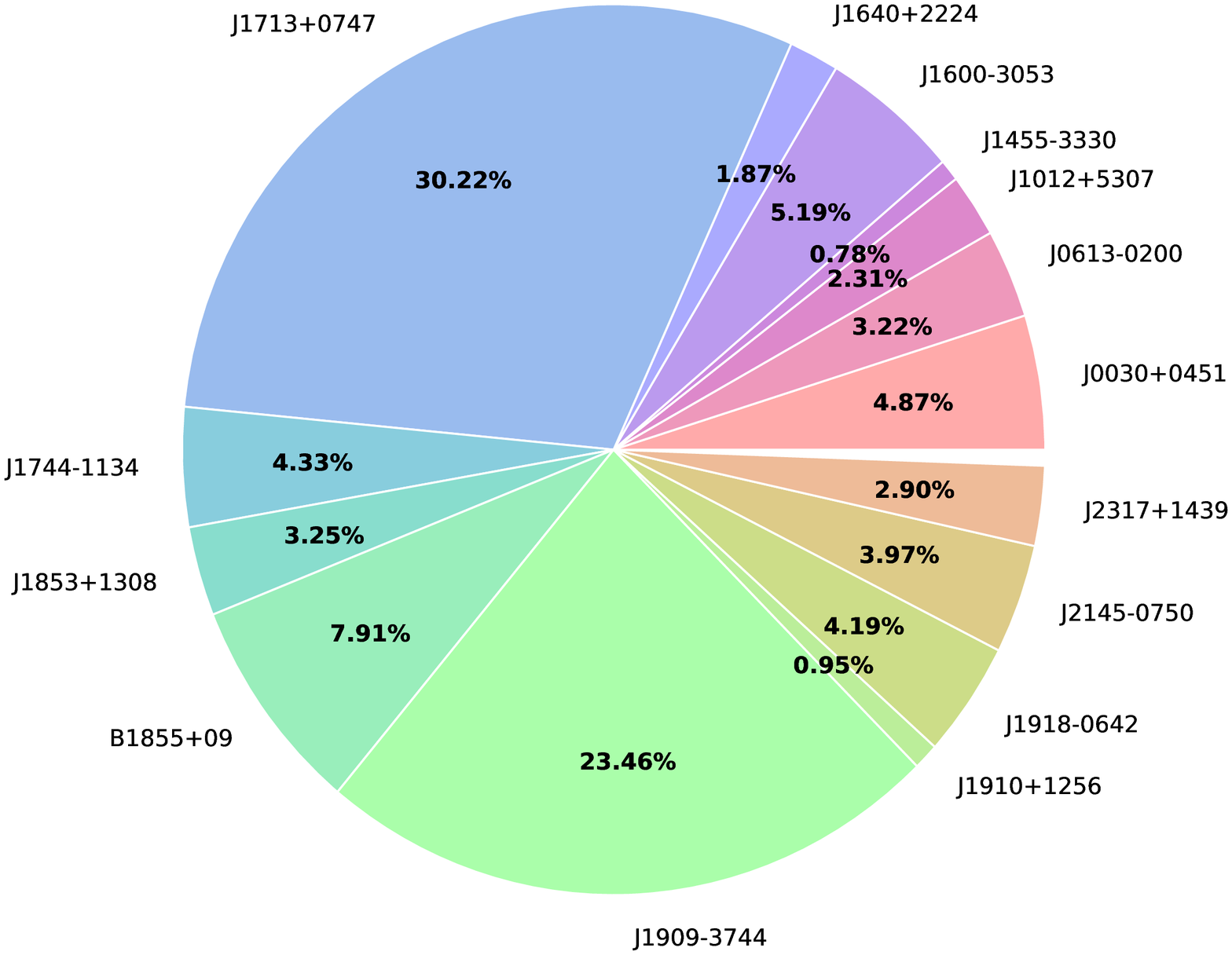}&
        \includegraphics[width=0.35\textwidth]{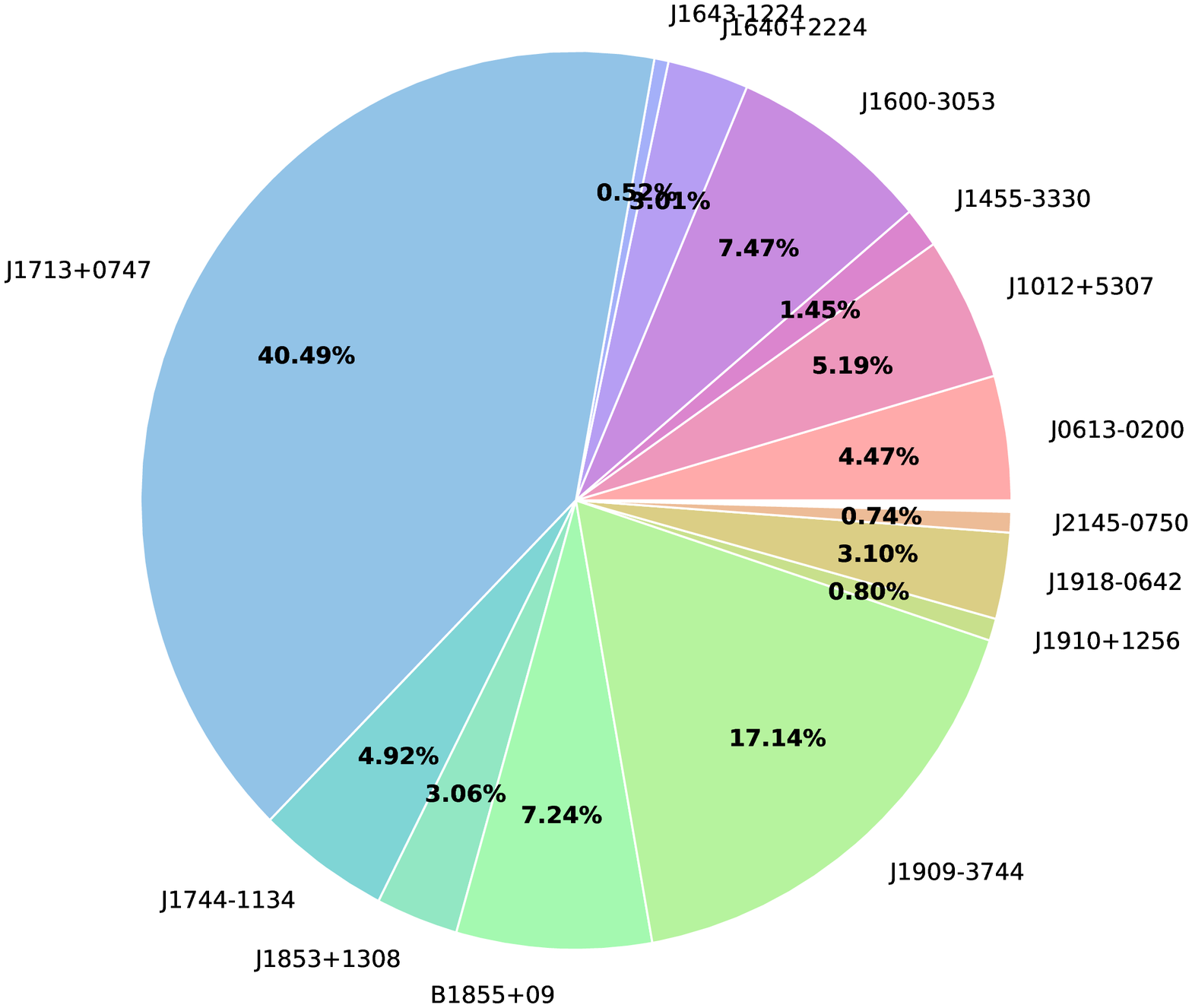}\\
        \bottomrule
        \end{tabular}
          \label{fig:perc_nf}
    \end{center}
\end{table}

\begin{table}
  \begin{center}
   \caption{Average time allocation $t_{j,\rm{frac}}$ for a noise floor drawn at random  Color version available online.} \begin{tabular}{|>{\centering\arraybackslash}m{0.16\textwidth}|>{\centering\arraybackslash}m{0.37\textwidth}|>{\centering\arraybackslash}m{0.37\textwidth}|}
        \toprule
        Time Allocation & Isotropic & GW from Virgo\\
        ($T_{\rm{Tot}}/T_{\rm{Tot0}}$) & & \\
        \midrule
        1x &
        \includegraphics[width=0.35\textwidth]{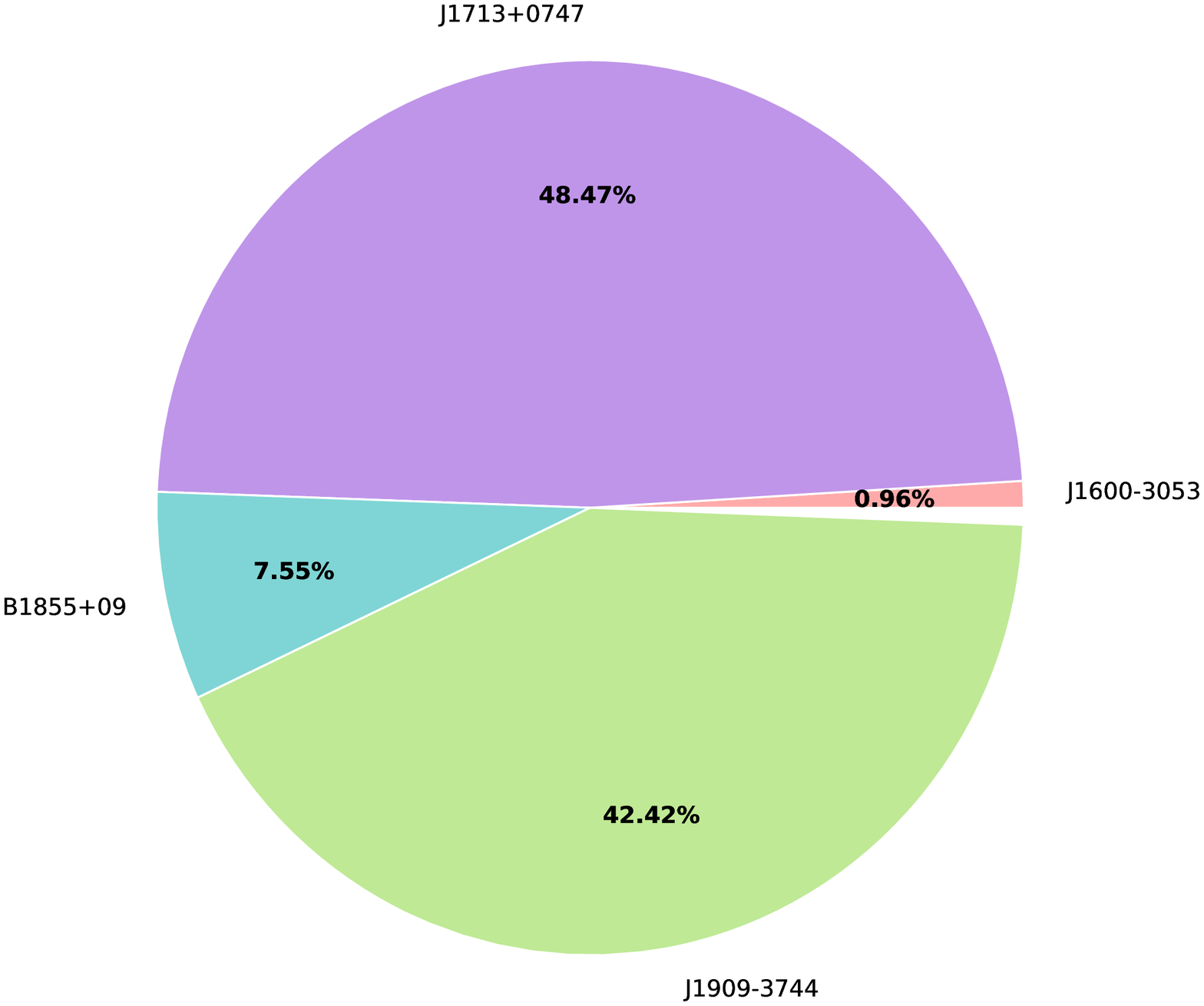} &
        \includegraphics[width=0.35\textwidth]{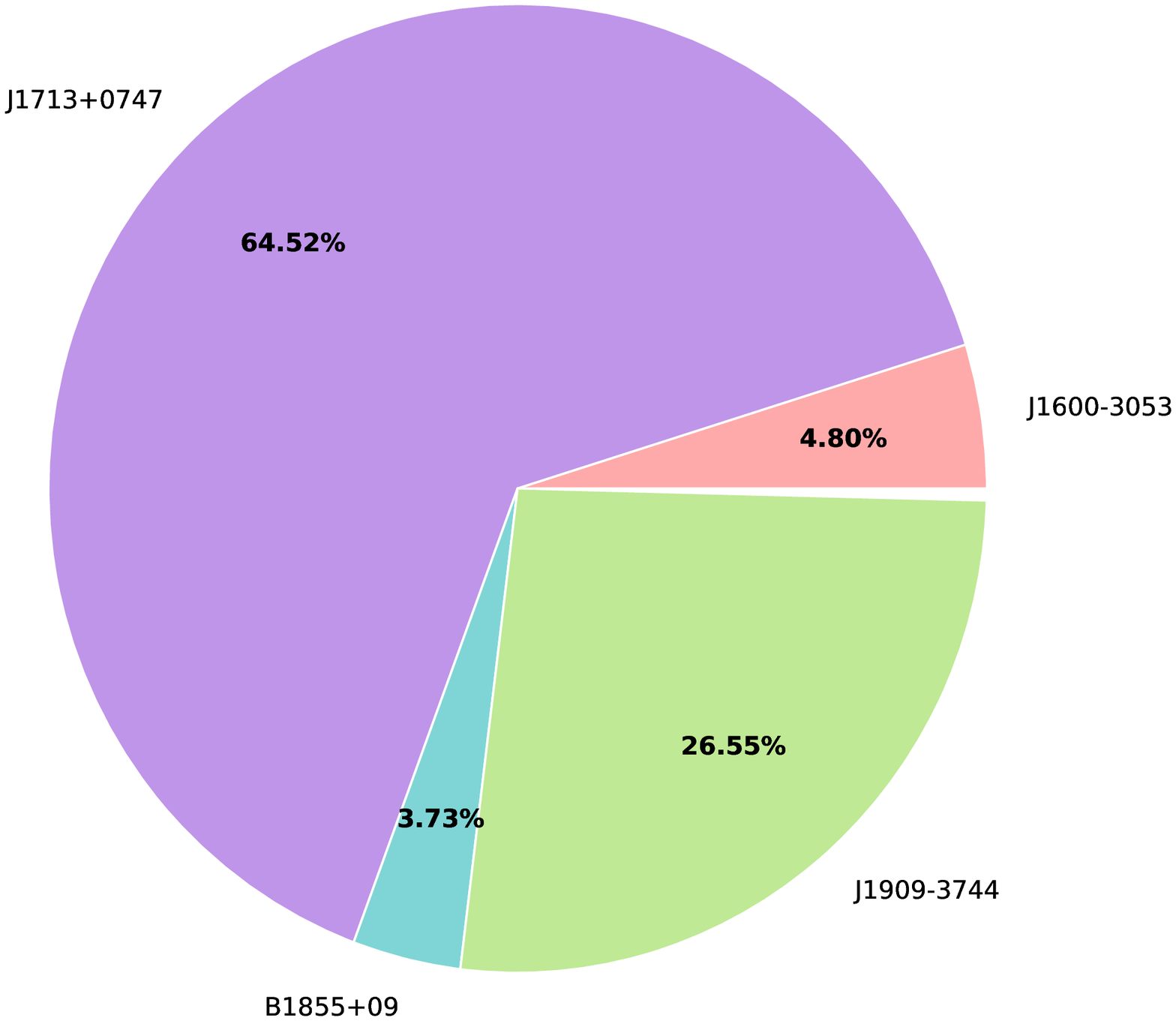} \\
        \midrule
        5x &
        \includegraphics[width=0.35\textwidth]{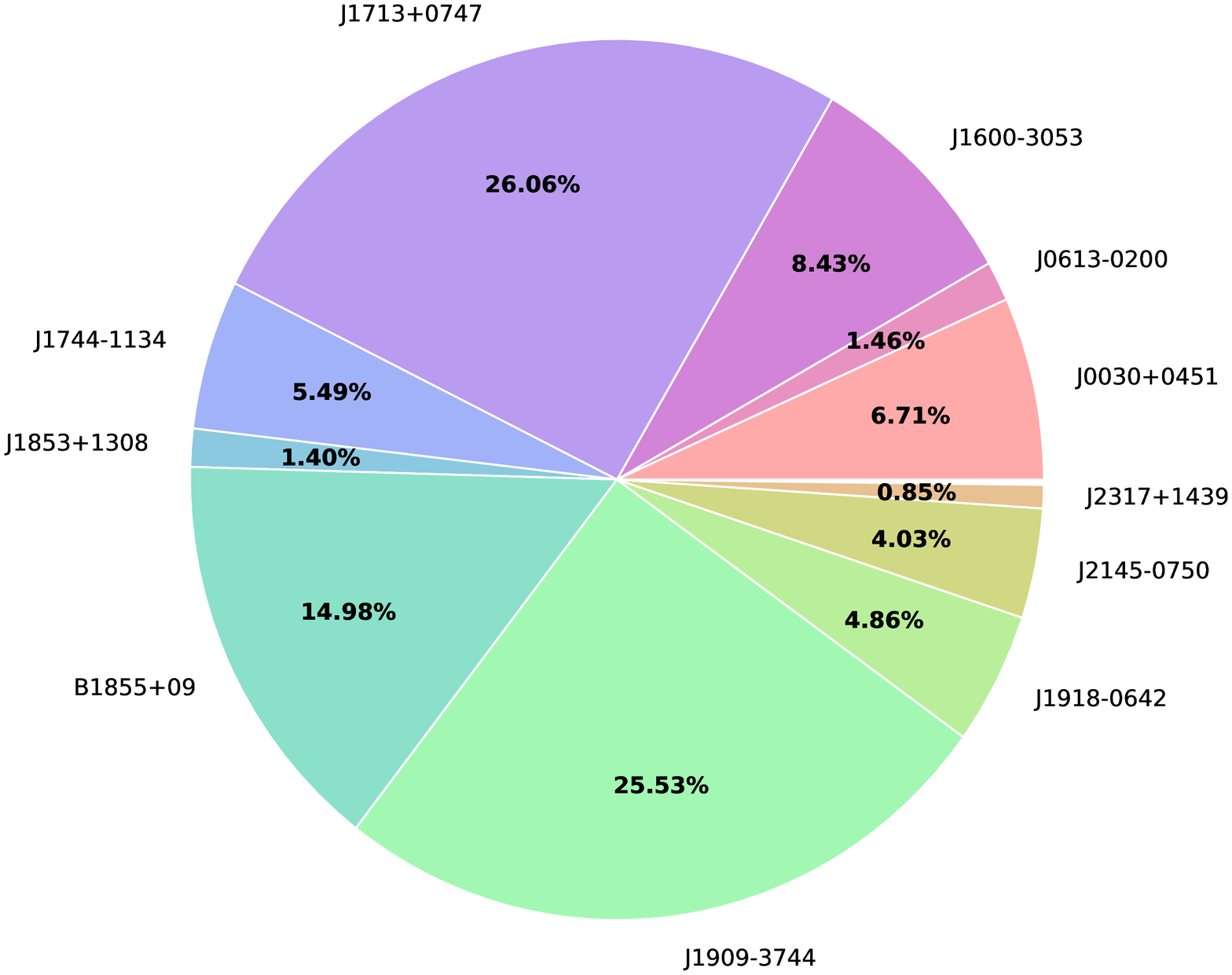} &
        \includegraphics[width=0.35\textwidth]{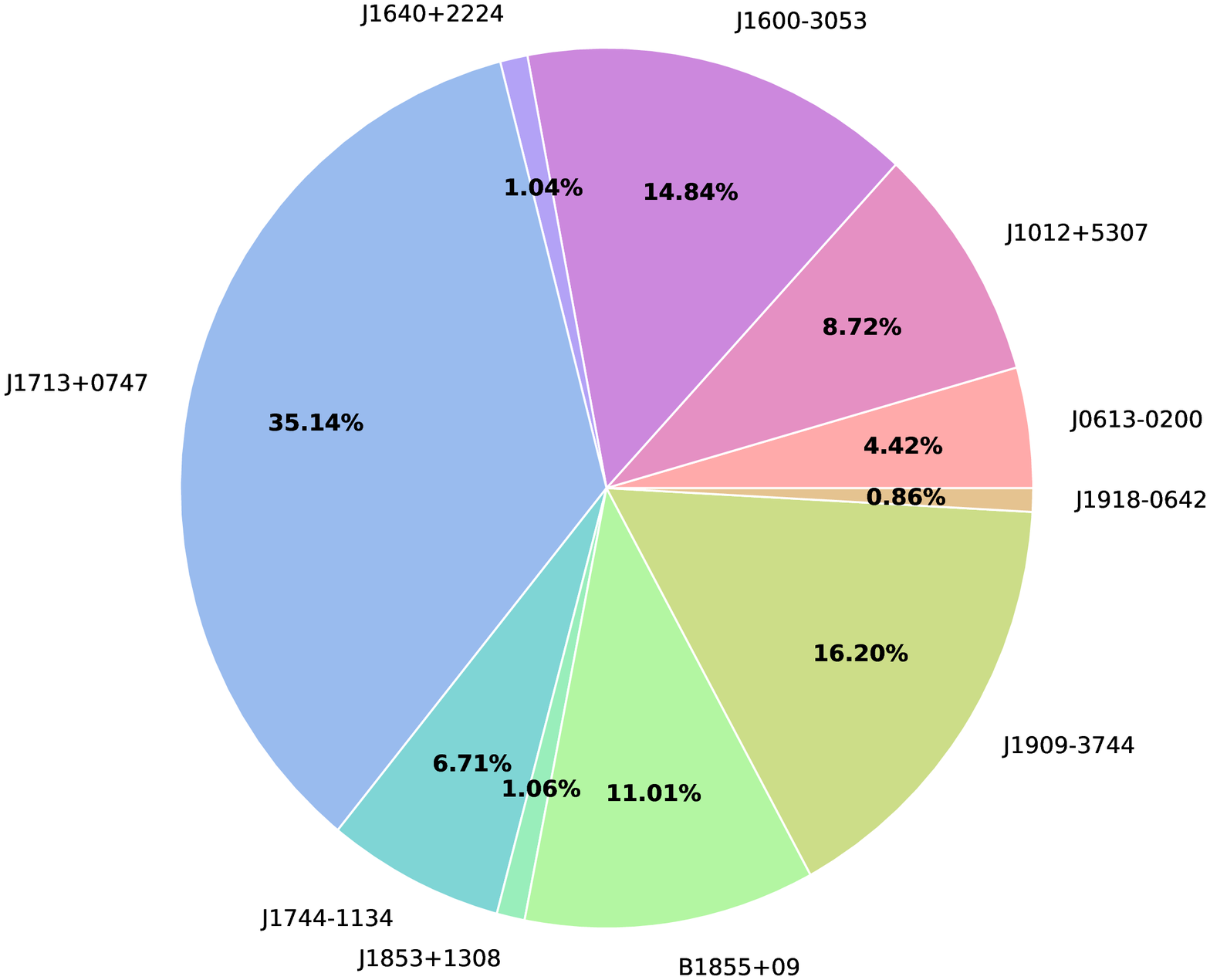} \\
        \midrule
        10x &
        \includegraphics[width=0.35\textwidth]{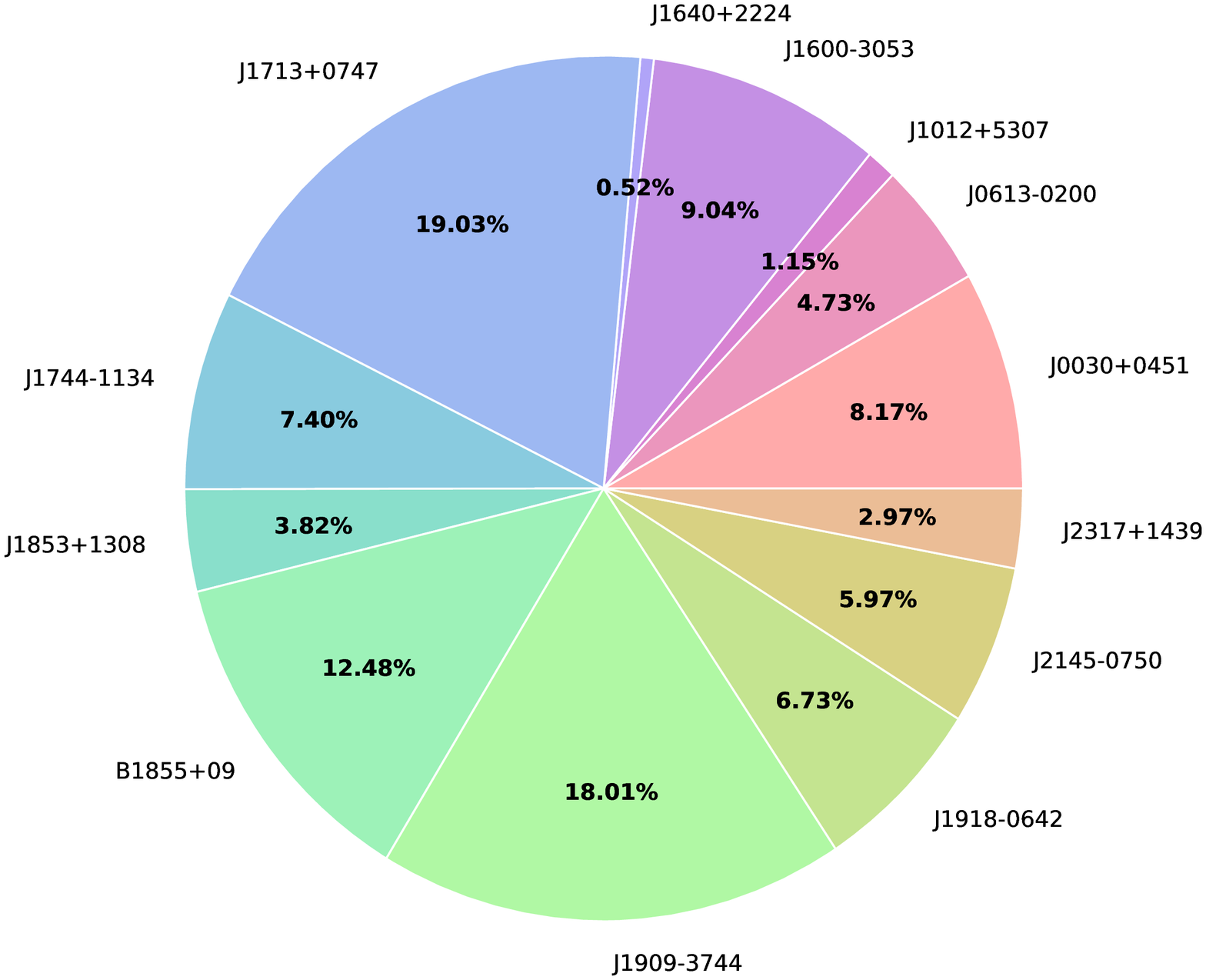}&
        \includegraphics[width=0.35\textwidth]{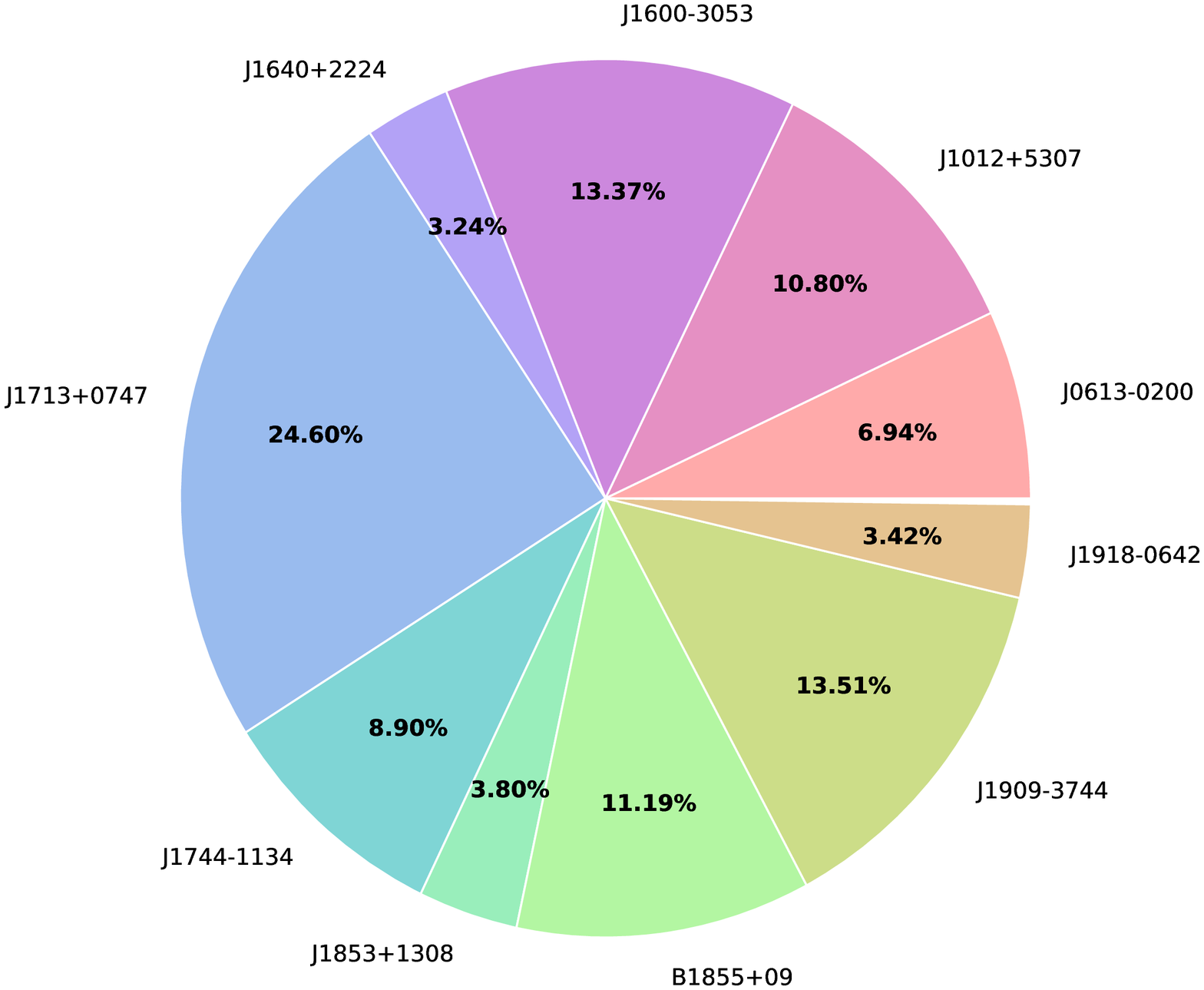}\\
        \bottomrule
        \end{tabular}
          \label{fig:rand_nf}
    \end{center}
\end{table}

\begin{table}[h]
\caption{The factor by which the sensitivity of the PTA increases with the optimized timing allocation.}
\begin{center}
    \begin{tabular}{|cc|c|c|}
    \toprule
    &&\multicolumn{2}{|c|}{Optimized Sensitivity/Reference Sensitivity}\\
    \hline
    Total Allocation time&Noise Floor Model&Isotropic&Virgo\\
    \hline\hline
    \multirow{3}{*}{Current Allocation Time} & 10 ns & 8.99 & 10.13\\
    & 80\% & 1.56 & 1.63\\
    & Random & 4.09 & 4.46\\
    \hline
    \multirow{3}{*}{Current Allocation x5} & 10 ns & 2.99 & 2.89\\
    & 80\% & 1.11 & 1.12\\
    & Random & 1.81 & 1.83\\
    \hline
    \multirow{3}{*}{Current Allocation x10} & 10 ns & 2.04 & 2.07\\
    & 80\% & 1.05 &1.06\\
    & Random & 1.53 & 1.53\\
    \bottomrule
    \end{tabular}
\end{center}
\label{tab:prim_results}
\end{table}

\section{Discussion}\label{sec:disc}
 We examined the time distribution of pulsars to maximize the volume sensitivity of a PTA to single sources for several noise floor conditions and total allocation times.  Table \ref{tab:prim_results} presents the improvement factor in each of these scenarios.  We again note that for larger time allocations than the current campaign, the reference PTA is one which is observed equally among all pulsars for the longer time, i.e., not the current data.

Several patterns emerge from the results.  First, optimization increases the overall volume sensitivity in all cases.  In comparing the different noise models, the largest gain occurs with the optimistically low constant noise floor.  The gain is due to the best pulsars (J1713+0747, J1909-3744) being sufficiently far away from the noise floor to improve with more allocation time.  We find that the improvement is roughly consistent in the case where a source is in the Virgo direction.  This is primarily due to the fact there is not a good pulsar directly in that part of the sky to benefit from more time.  As seen in Table \ref{tab:PulsarInfo}, only two of the six pulsars with a $(1-\hat{k}_{\rm{Virgo}} \cdot \hat{n})^2$ above 1.5 have an RMS below 200 ns.

We can also consider the allocation results from each noise floor case individually.

{\it Constant Noise Floor}: The constant noise floor case is an optimistic situation where we do not encounter the noise floor quickly.  In this scenario, the result is for the timing to focus on the best two pulsars, J1713+0747 and J1909-3744.  However, when we have more total allocation time, we see the effect of the noise floor by time being diverted to more pulsars.  This presents the general relationship with allocation time: the best pulsar is favored until its timing allocation causes the RMS to reach the noise floor, then time begins being diverted to the second best, and so on.  The case with Virgo shows the directional effect as both J1713+0747 and J1600-3053 receive more time due to their large $(1-\hat{k}_{\rm{Virgo}}\cdot \hat{n})^2$ factor relative to the rest of the pulsars.  Conversely, J1909-3744 and B1855+09 receive less allocation for the same reason.

{\it 80\% Noise Floor Value for Each Pulsar}: Switching to a more pessimistic scenario, we see the noise floor being reached quickly for the best pulsars.  In this case, many more pulsars are part of the optimized array.  While the optimization still favors focusing on the two best pulsars, as more allocation time becomes available the long term drift is towards a more even distribution.  However, even with a factor of 10 increase, the two best pulsars receive over half the available time.  Here it is easiest to see the effect of being in a favorable location in relation to Virgo, as the three best placed pulsars (J1012+5307, J1455-3330, and J1640+2224) receive substantial increases in their allocation by as much as a factor of 2.  Even J1713+0747 receives an increase $\sim 20\%-30\%$.

{\it Noise Floor Calculated at Random}: This scenario predictably
falls between the two cases, with sensitivity increases roughly equal
to the geometric mean of the first two models.  However, for larger
time allocations, J1713+0747 and J1909-3744 both receive less time
than the other two noise models.  To understand this note that any
individual simulation may assign these pulsars a relatively high noise
floor compared to the other pulsars.  This is a case not explored
previously, in which J1713+0747 and J1909-3744 always had the best (or
equal) noise floor value.  

\begin{figure}
\includegraphics[scale=0.8]{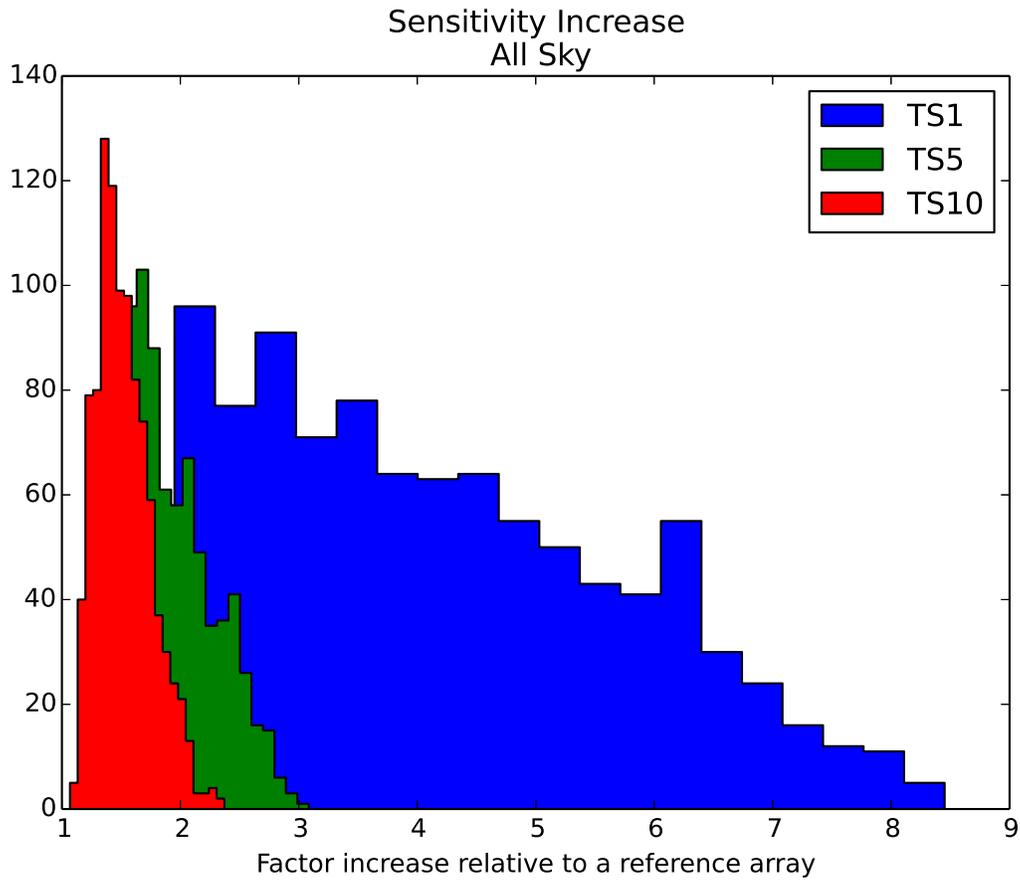}
\caption{Distribution of sensitivity increase for each trial of the
   case with noise floor being drawn from random and optimized for the
 Virgo cluster direction.  Color version available online}
\label{fig:allsky_var} 
\end{figure}

\begin{figure}
\includegraphics[scale=0.8]{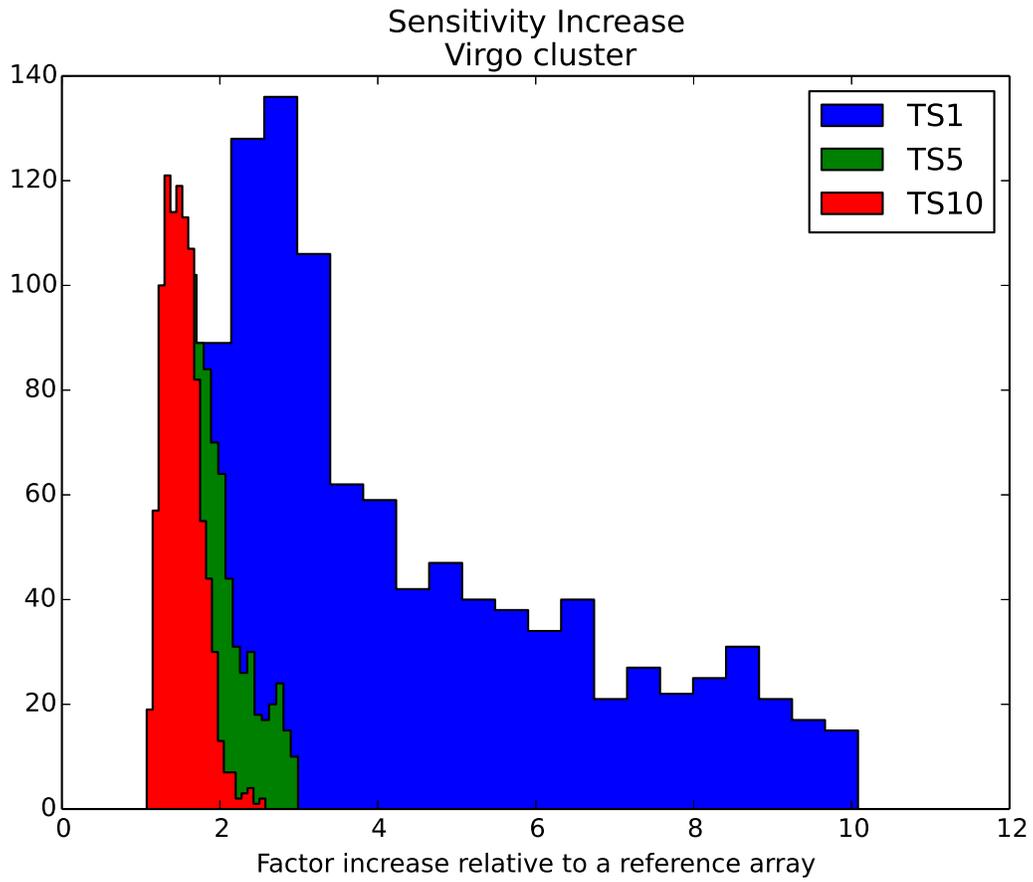}
\caption{Distribution of sensitivity increase for each trial of the
   case with noise floor being drawn from random and optimized for the
 Virgo cluster direction.  Color version available online}
\label{fig:virgo_var} 
\end{figure}

The robustness of the optimization scheme is visualized in
  Figures \ref{fig:allsky_var},\ref{fig:virgo_var}.  This displays the
  distribution of
  sensitivity increases for each of the 1000 realizations.  As
  expected, the variation decreases as the available time for
  allocation increases, due to more pulsars reaching the noise floor
  and time allocation approaching an even distribution.  Despite this,
  it is always the case there is improvement over the reference array.


\section{Conclusion}
The optimization of a PTA's sensitivity is very important to improve the chance of detecting a GW. There are many factors in determining the sensitivity of an array. This work aims to add to the overall effort by considering an often overlooked signal: the single source and/or burst.  The primary result is that one should spend more time than the current setup on pulsars that are well-timed, particularly J1713+0747 and J1909-3744.  

A fundamental input necessary to make these results more robust is the actual noise floors for each pulsar.  Dedicated campaigns to determine the floor would be valuable.  In June of 2013, one such campaign was done on J1713+0747 where it was timed continuously for 24 hours by several telescopes within the IPTA \citep{Dolch14}.  We expect to use the results of this study to more accurately model the noise floor of this pulsar, which is the most important ingredient in our optimization scheme.  Another input is the description of the time allocated to each pulsar used in generating the input RMS values ($T_{j0}$).  These values are not exactly even among the pulsars.  Calculating this is non-trivial as the data is taken at two different sites and the time spent on each is not an apples-to-apples comparison.  

This result should be placed in the context of optimization studies performed for the stochastic background (\cite{Siemens13},\cite{Lee12}).  Here, the statistic optimized includes the Hellings and Downs correlation coefficient \citep{hellings:gw} that scales according to the number of pulsars.  Therefore, their statistic is improved with more pulsars unlike the one optimized here.  We stress that the approaches differ due to the difference of the signal being considered.

There is also the issue of confirmation that having a correlated response across many pulsars
provides. It is true that for the first detection, this correlation will be vital to distinguish from other
noise effects, such as a glitch \citep{Cognard04}.  How an individual 
detection pipeline handles potential false positives is beyond the scope of this work.   
However, once the detection has been made and well justified, future observations
will be less constricted by this argument. In this way, one can consider these results as pertaining
to the era of gravitational wave astronomy. That said, it is important to note that in all cases, at
least two pulsars are considered in the final optimization. This contrasts with the result of just
white noise, where all time is devoted to J1713+0747. Indeed, for many of the schemes involving
the 80\% noise floor, the number of pulsars is $\sim 10$.

These considerations lead us away from the statement that these optimization results alone should dictate the timing allocation of PTAs, specifically that the allocation time should be completely devoted to J1713+0747 and J1909-3744.  Rather, we see this work as a counterbalance to the claim that more pulsars are always better.  As such, our recommendation would be to devote a certain fraction of the time evenly among all the pulsars, and use the remaining time to observe the best pulsars more, either per observation or by increasing the cadence.  At the start, it should simply be J1713+0747 and J1909-3744. Which pulsars to favor beyond this depends upon the underlying noise floor value, though B1855+09 is the likely third choice.

Focusing on Virgo (or any other particular direction) will also affect
which handful of pulsars one decides to spend more time on.  Scaling
the RMS by $1/(1-\hat{k} \cdot \hat{n})$ provides the equivalent
problem in an isotropic case.  Therefore, if one is to focus the
sensitivity of the array in a particular direction, pulsars within
that direction will be more favored for time allocation.  As
\citet{Simon14} has demonstrated, a single source from the local
universe is more likely in the direction of galaxy clusters.  Pulsars
near the direction of these clusters should be favored.  This also
demonstrates the importance of focusing searches for new pulsars in
the region around these clusters.  For example, if a pulsar with 450
ns RMS is discovered 60$^{\circ}$ away from Virgo, this provides the
same sensitivity as a pulsar with 150 ns that is 120$^{\circ}$ away.
Note again this work only applies to sources separated
  enough from the pulsar so the optimization neglects the pulsar
  term.

Our optimization shows that it is possible to change the allocation time of NANOGrav and increase the volume of space for detectable single source signals by several factors.  Future efforts will be directed at improving the accuracy of these results and considering broader answers to the questions outlined in \S 1.  This work aims to provide a broad range of noise floor values to reflect our ignorance of the true result.  An important improvement will be to continually increase the accuracy of our noise floor estimates.  The results of the global 24-hour timing campaign of J1713+0747 will be a valuable step in this direction.  Further, this work only considers the ability to detect single sources.  However, to perform astronomy, these waves must be characterized.  This will be the focus of \cite{Koop}.


\begin{thebibliography}{}

\bibitem[{Backer} {et~al.}(1982){Backer}, {Kulkarni}, {Heiles}, {Davis}, and
  {Goss}]{Backer82}
{Backer}, D.~C., {Kulkarni}, S.~R., {Heiles}, C., {Davis}, M.~M., \& {Goss},
  W.~M. 1982, \nat, 300, 615

\bibitem[{Burt}, {Lommen}, \& {Finn}(2011){Burt}, {Lommen}, and {Finn}]{Burt11}
{Burt}, B.~J., {Lommen}, A.~N., \& {Finn}, L.~S. 2011, \apj, 730, 17

\bibitem[{Cognard} \& {Backer}(2004){Cognard} and {Backer}]{Cognard04}
{Cognard}, I., \& {Backer}, D.~C. 2004, \apjl, 612, L125--L127

\bibitem[Cordes(2013)Cordes]{Cordes13}
Cordes, J.~M. 2013, Classical and Quantum Gravity, 30, 224002

\bibitem[{Cordes} \& {Jenet}(2012){Cordes} and {Jenet}]{Cordes12}
{Cordes}, J.~M., \& {Jenet}, F.~A. 2012, \apj, 752, 54

\bibitem[{Cordes} \& {Lazio}(2002){Cordes} and {Lazio}]{Cordes02}
{Cordes}, J.~M., \& {Lazio}, T.~J. 2002, astro-ph/0207156

\bibitem[{Cutler} {et~al.}(2014){Cutler}, {Burke-Spolaor}, {Vallisneri},
  {Lazio}, and {Majid}]{Cutler13}
{Cutler}, C., {Burke-Spolaor}, S., {Vallisneri}, M., {Lazio}, J., \& {Majid},
  W. 2014, \prd, 89, 042003

\bibitem[{Demorest} {et~al.}(2013){Demorest}, {Ferdman}, {Gonzalez}, {Nice},
  {Ransom}, {Stairs}, {Arzoumanian}, {Brazier}, {Burke-Spolaor}, {Chamberlin},
  {Cordes}, {Ellis}, {Finn}, {Freire}, {Giampanis}, {Jenet}, {Kaspi}, {Lazio},
  {Lommen}, {McLaughlin}, {Palliyaguru}, {Perrodin}, {Shannon}, {Siemens},
  {Stinebring}, {Swiggum}, and {Zhu}]{Demorest13}
{Demorest}, P.~B., {Ferdman}, R.~D., {Gonzalez}, M.~E., {Nice}, D., {Ransom},
  S., {Stairs}, I.~H., {Arzoumanian}, Z., {Brazier}, A., {Burke-Spolaor}, S.,
  {Chamberlin}, S.~J., {Cordes}, J.~M., {Ellis}, J., {Finn}, L.~S., {Freire},
  P., {Giampanis}, S., {Jenet}, F., {Kaspi}, V.~M., {Lazio}, J., {Lommen},
  A.~N., {McLaughlin}, M., {Palliyaguru}, N., {Perrodin}, D., {Shannon}, R.~M.,
  {Siemens}, X., {Stinebring}, D., {Swiggum}, J., \& {Zhu}, W.~W. 2013, \apj,
  762, 94

\bibitem[{Dolch et al.}(2014){Dolch et al.}]{Dolch14}
{Dolch et al.} 2014, In Preparation

\bibitem[{Ferdman} {et~al.}(2010){Ferdman}, {van Haasteren}, {Bassa}, {Burgay},
  {Cognard}, {Corongiu}, {D'Amico}, {Desvignes}, {Hessels}, {Janssen},
  {Jessner}, {Jordan}, {Karuppusamy}, {Keane}, {Kramer}, {Lazaridis}, {Levin},
  {Lyne}, {Pilia}, {Possenti}, {Purver}, {Stappers}, {Sanidas}, {Smits}, and
  {Theureau}]{Ferdman10}
{Ferdman}, R.~D., {van Haasteren}, R., {Bassa}, C.~G., {Burgay}, M., {Cognard},
  I., {Corongiu}, A., {D'Amico}, N., {Desvignes}, G., {Hessels}, J.~W.~T.,
  {Janssen}, G.~H., {Jessner}, A., {Jordan}, C., {Karuppusamy}, R., {Keane},
  E.~F., {Kramer}, M., {Lazaridis}, K., {Levin}, Y., {Lyne}, A.~G., {Pilia},
  M., {Possenti}, A., {Purver}, M., {Stappers}, B., {Sanidas}, S., {Smits}, R.,
  \& {Theureau}, G. 2010, Classical and Quantum Gravity, 27, 084014

\bibitem[{Finn} \& {Lommen}(2010){Finn} and {Lommen}]{Finn10}
{Finn}, L.~S., \& {Lommen}, A.~N. 2010, \apj, 718, 1400

\bibitem[{Foster} \& {Backer}(1990){Foster} and {Backer}]{Foster90}
{Foster}, R.~S., \& {Backer}, D.~C. 1990, \apj, 361, 300

\bibitem[{Handzo} {et~al.}(2014){Handzo}, {Christy}, {Lommen}, and
  {Perrodin}]{Handzo}
{Handzo}, E., {Christy}, B., {Lommen}, A., \& {Perrodin}, D. 2014, Submitted to
  Astrophyiscal Journal

\bibitem[{Hellings} \& {Downs}(1983){Hellings} and {Downs}]{hellings:gw}
{Hellings}, R.~W., \& {Downs}, G.~S. 1983, \apjl, 265, L39--L42

\bibitem[{Hemberger} \& {Stinebring}(2008){Hemberger} and
  {Stinebring}]{Hemberger08}
{Hemberger}, D.~A., \& {Stinebring}, D.~R. 2008, \apjl, 674, L37

\bibitem[{Hobbs}, {Lyne}, \& {Kramer}(2006){Hobbs}, {Lyne}, and
  {Kramer}]{Hobbs06}
{Hobbs}, G., {Lyne}, A., \& {Kramer}, M. 2006, Chinese Journal of Astronomy and
  Astrophysics Supplement, 6, 169

\bibitem[{Hobbs} {et~al.}(2010){Hobbs}, {Archibald}, {Arzoumanian}, {Backer},
  {Bailes}, {Bhat}, {Burgay}, {Burke-Spolaor}, {Champion}, {Cognard}, {Coles},
  {Cordes}, {Demorest}, {Desvignes}, {Ferdman}, {Finn}, {Freire}, {Gonzalez},
  {Hessels}, {Hotan}, {Janssen}, {Jenet}, {Jessner}, {Jordan}, {Kaspi},
  {Kramer}, {Kondratiev}, {Lazio}, {Lazaridis}, {Lee}, {Levin}, {Lommen},
  {Lorimer}, {Lynch}, {Lyne}, {Manchester}, {McLaughlin}, {Nice}, {Oslowski},
  {Pilia}, {Possenti}, {Purver}, {Ransom}, {Reynolds}, {Sanidas}, {Sarkissian},
  {Sesana}, {Shannon}, {Siemens}, {Stairs}, {Stappers}, {Stinebring},
  {Theureau}, {van Haasteren}, {van Straten}, {Verbiest}, {Yardley}, and
  {You}]{Hobbs10}
{Hobbs}, G., {Archibald}, A., {Arzoumanian}, Z., {Backer}, D., {Bailes}, M.,
  {Bhat}, N.~D.~R., {Burgay}, M., {Burke-Spolaor}, S., {Champion}, D.,
  {Cognard}, I., {Coles}, W., {Cordes}, J., {Demorest}, P., {Desvignes}, G.,
  {Ferdman}, R.~D., {Finn}, L., {Freire}, P., {Gonzalez}, M., {Hessels}, J.,
  {Hotan}, A., {Janssen}, G., {Jenet}, F., {Jessner}, A., {Jordan}, C.,
  {Kaspi}, V., {Kramer}, M., {Kondratiev}, V., {Lazio}, J., {Lazaridis}, K.,
  {Lee}, K.~J., {Levin}, Y., {Lommen}, A., {Lorimer}, D., {Lynch}, R., {Lyne},
  A., {Manchester}, R., {McLaughlin}, M., {Nice}, D., {Oslowski}, S., {Pilia},
  M., {Possenti}, A., {Purver}, M., {Ransom}, S., {Reynolds}, J., {Sanidas},
  S., {Sarkissian}, J., {Sesana}, A., {Shannon}, R., {Siemens}, X., {Stairs},
  I., {Stappers}, B., {Stinebring}, D., {Theureau}, G., {van Haasteren}, R.,
  {van Straten}, W., {Verbiest}, J.~P.~W., {Yardley}, D.~R.~B., \& {You}, X.~P.
  2010, Classical and Quantum Gravity, 27, 084013

\bibitem[{Hulse} \& {Taylor}(1975){Hulse} and {Taylor}]{Hulse75}
{Hulse}, R.~A., \& {Taylor}, J.~H. 1975, \apjl, 195, L51

\bibitem[{Jaffe} \& {Backer}(2003){Jaffe} and {Backer}]{jb03}
{Jaffe}, A.~H., \& {Backer}, D.~C. 2003, \apj, 583, 616--631

\bibitem[{Jenet} {et~al.}(2009){Jenet}, {Finn}, {Lazio}, {Lommen},
  {McLaughlin}, {Stairs}, {Stinebring}, {Verbiest}, {Archibald}, {Arzoumanian},
  {Backer}, {Cordes}, {Demorest}, {Ferdman}, {Freire}, {Gonzalez}, {Kaspi},
  {Kondratiev}, {Lorimer}, {Lynch}, {Nice}, {Ransom}, {Shannon}, and
  {Siemens}]{Jenet09}
{Jenet}, F., {Finn}, L.~S., {Lazio}, J., {Lommen}, A., {McLaughlin}, M.,
  {Stairs}, I., {Stinebring}, D., {Verbiest}, J., {Archibald}, A.,
  {Arzoumanian}, Z., {Backer}, D., {Cordes}, J., {Demorest}, P., {Ferdman}, R.,
  {Freire}, P., {Gonzalez}, M., {Kaspi}, V., {Kondratiev}, V., {Lorimer}, D.,
  {Lynch}, R., {Nice}, D., {Ransom}, S., {Shannon}, R., \& {Siemens}, X. 2009,
  ArXiv e-prints

\bibitem[{Jenet} {et~al.}(2004){Jenet}, {Lommen}, {Larson}, and {Wen}]{Jenet04}
{Jenet}, F.~A., {Lommen}, A., {Larson}, S.~L., \& {Wen}, L. 2004, \apj, 606,
  799

\bibitem[{Jenet} {et~al.}(2005){Jenet}, {Hobbs}, {Lee}, and
  {Manchester}]{Jenet05detect}
{Jenet}, F.~A., {Hobbs}, G.~B., {Lee}, K.~J., \& {Manchester}, R.~N. 2005,
  \apjl, 625, L123

\bibitem[Koop \& Finn(2014)Koop and Finn]{Koop}
Koop, M., \& Finn, L.~S. 2014, in preparation

\bibitem[{Kramer} {et~al.}(2006){Kramer}, {Stairs}, {Manchester}, {McLaughlin},
  {Lyne}, {Ferdman}, {Burgay}, {Lorimer}, {Possenti}, {D'Amico}, {Sarkissian},
  {Hobbs}, {Reynolds}, {Freire}, and {Camilo}]{Kramer06}
{Kramer}, M., {Stairs}, I.~H., {Manchester}, R.~N., {McLaughlin}, M.~A.,
  {Lyne}, A.~G., {Ferdman}, R.~D., {Burgay}, M., {Lorimer}, D.~R., {Possenti},
  A., {D'Amico}, N., {Sarkissian}, J.~M., {Hobbs}, G.~B., {Reynolds}, J.~E.,
  {Freire}, P.~C.~C., \& {Camilo}, F. 2006, Science, 314, 97--102

\bibitem[{Lee} {et~al.}(2012){Lee}, {Bassa}, {Janssen}, {Karuppusamy},
  {Kramer}, {Smits}, and {Stappers}]{Lee12}
{Lee}, K.~J., {Bassa}, C.~G., {Janssen}, G.~H., {Karuppusamy}, R., {Kramer},
  M., {Smits}, R., \& {Stappers}, B.~W. 2012, \mnras, 423, 2642--2655

\bibitem[{Manchester} {et~al.}(2013){Manchester}, {Hobbs}, {Bailes}, {Coles},
  {van Straten}, {Keith}, {Shannon}, {Bhat}, {Brown}, {Burke-Spolaor},
  {Champion}, {Chaudhary}, {Edwards}, {Hampson}, {Hotan}, {Jameson}, {Jenet},
  {Kesteven}, {Khoo}, {Kocz}, {Maciesiak}, {Oslowski}, {Ravi}, {Reynolds},
  {Sarkissian}, {Verbiest}, {Wen}, {Wilson}, {Yardley}, {Yan}, and
  {You}]{Manchester13}
{Manchester}, R.~N., {Hobbs}, G., {Bailes}, M., {Coles}, W.~A., {van Straten},
  W., {Keith}, M.~J., {Shannon}, R.~M., {Bhat}, N.~D.~R., {Brown}, A.,
  {Burke-Spolaor}, S.~G., {Champion}, D.~J., {Chaudhary}, A., {Edwards}, R.~T.,
  {Hampson}, G., {Hotan}, A.~W., {Jameson}, A., {Jenet}, F.~A., {Kesteven},
  M.~J., {Khoo}, J., {Kocz}, J., {Maciesiak}, K., {Oslowski}, S., {Ravi}, V.,
  {Reynolds}, J.~R., {Sarkissian}, J.~M., {Verbiest}, J.~P.~W., {Wen}, Z.~L.,
  {Wilson}, W.~E., {Yardley}, D., {Yan}, W.~M., \& {You}, X.~P. 2013, 
  Publications of the Astronomical Society of Australia, 30, 17

\bibitem[{Os{\l}owski} {et~al.}(2011){Os{\l}owski}, {van Straten}, {Hobbs},
  {Bailes}, and {Demorest}]{Oslowski11}
{Os{\l}owski}, S., {van Straten}, W., {Hobbs}, G.~B., {Bailes}, M., \&
  {Demorest}, P. 2011, \mnras, 418, 1258

\bibitem[Perrodin {et~al.}(2014)Perrodin, Jenet, Lommen, Finn, Demorest,
  Ferdman, Gonzalez, Nice, Ransom, and Stairs]{Perrodin}
Perrodin, D., Jenet, F., Lommen, A.~N., Finn, L.~S., Demorest, P.~B., Ferdman,
  R.~D., Gonzalez, M.~E., Nice, D., Ransom, S., \& Stairs, I.~H. 2014,
  Submitted to Astrophysical Journal

\bibitem[{Ravi} {et~al.}(2014){Ravi}, {Wyithe}, {Shannon}, {Hobbs}, and
  {Manchester}]{Ravi14}
{Ravi}, V., {Wyithe}, J.~S.~B., {Shannon}, R.~M., {Hobbs}, G., \& {Manchester},
  R.~N. 2014, ArXiv e-prints

\bibitem[{Sesana}(2013){Sesana}]{Sesana13}
{Sesana}, A. 2013, \mnras, 433, L1--L5

\bibitem[{Sesana}, {Vecchio}, \& {Colacino}(2008){Sesana}, {Vecchio}, and
  {Colacino}]{svc08}
{Sesana}, A., {Vecchio}, A., \& {Colacino}, C.~N. 2008, \mnras, 390, 192--209

\bibitem[{Shannon} \& {Cordes}(2010){Shannon} and {Cordes}]{Shannon10}
{Shannon}, R.~M., \& {Cordes}, J.~M. 2010, \apj, 725, 1607

\bibitem[{Siemens}, {Mandic}, \& {Creighton}(2007){Siemens}, {Mandic}, and
  {Creighton}]{Siemens07}
{Siemens}, X., {Mandic}, V., \& {Creighton}, J. 2007, Physical Review Letters,
  98, 111101

\bibitem[{Siemens} {et~al.}(2013){Siemens}, {Ellis}, {Jenet}, and
  {Romano}]{Siemens13}
{Siemens}, X., {Ellis}, J., {Jenet}, F., \& {Romano}, J.~D. 2013, ArXiv
  e-prints

\bibitem[{Simon} {et~al.}(2014){Simon}, {Polin}, {Lommen}, {Stappers}, {Finn},
  {Jenet}, and {Christy}]{Simon14}
{Simon}, J., {Polin}, A., {Lommen}, A., {Stappers}, B., {Finn}, L.~S., {Jenet},
  F.~A., \& {Christy}, B. 2014, \apj, 784, 60

\bibitem[{Taylor}(1992){Taylor}]{Taylor92}
{Taylor}, J.~H. 1992, Philosophical Transactions of the Royal Society of
  London, 341, 117-134 (1992), 341, 117

\bibitem[{You} {et~al.}(2007){You}, {Hobbs}, {Coles}, {Manchester}, {Edwards},
  {Bailes}, {Sarkissian}, {Verbiest}, {van Straten}, {Hotan}, {Ord}, {Jenet},
  {Bhat}, and {Teoh}]{You07}
{You}, X.~P., {Hobbs}, G., {Coles}, W.~A., {Manchester}, R.~N., {Edwards}, R.,
  {Bailes}, M., {Sarkissian}, J., {Verbiest}, J.~P.~W., {van Straten}, W.,
  {Hotan}, A., {Ord}, S., {Jenet}, F., {Bhat}, N.~D.~R., \& {Teoh}, A. 2007,
  MNRAS, 378, 493

\end{thebibliography}
\end{document}